\documentstyle[11pt,epsf]{article}
\newcommand{\str}{\quad \Rightarrow \quad}

\newcommand{\fet}[1]{\mbox{\boldmath $#1$}}

\newcommand{\beq}{\begin{equation}}
\newcommand{\eeq}{\end{equation}}
\newcommand{\beqa}{\begin{eqnarray}}
\newcommand{\eeqa}{\end{eqnarray}}

\newcommand{\vs}{\vspace{-0.2cm}}
\newcommand{\pr}{\overrightarrow}
\newcommand{\lev}{\overleftarrow}

\begin{document}

%%{\bf DRAFT} 

\hfill KFA-IKP(TH)-1997-19

%%{\bf \today} 

\hfill nucl-th/9801nnn

\vspace{1in}

\begin{center}

{{\Large\bf Nuclear forces from chiral Lagrangians using the method\\[0.3em]
%using the method of 
of unitary transformation I~: Formalism}}

\end{center}

\vspace{.3in}

\begin{center}

{\large 
E. Epelbaoum,$^\ddagger$$^\dagger$\footnote{email: 
                           evgeni.epelbaum@hadron.tp2.ruhr-uni-bochum.de}
W. Gl\"ockle,$^\dagger$\footnote{email:
                           walter.gloeckle@hadron.tp2.ruhr-uni-bochum.de}
Ulf-G. Mei{\ss}ner$^\ddagger$\footnote{email: 
                           Ulf-G.Meissner@fz-juelich.de}}

\bigskip

$^\ddagger${\it Forschungszentrum J\"ulich, Institut f\"ur Kernphysik 
(Theorie)\\ D-52425 J\"ulich, Germany}

\bigskip

$^\dagger${\it Ruhr-Universit\"at Bochum, Institut f{\"u}r
  Theoretische Physik II\\ D-44870 Bochum, Germany}\\

\end{center}

\vspace{.9in}

\thispagestyle{empty}

\begin{abstract}
\noindent We construct the two-- and three--nucleon potential based
on the most general chiral effective pion--nucleon Lagrangian using
the method of unitary transformations. For that, we develop a power
counting scheme consistent with this projection formalism. In contrast
to previous results obtained in old--fashioned
 time--ordered perturbation theory, the
method employed leads to energy--independent potentials. We discuss in
detail the similarities and differences to the existing chiral
nucleon--nucleon potentials. We also show that to leading order in
the power counting, the three--nucleon forces vanish lending credit
to the result obtained by Weinberg using old--fashioned
time--ordered perturbation theory.  
\end{abstract}

\vfill

\pagebreak

%\end{document}

\section{Introduction}
\def\theequation{\arabic{section}.\arabic{equation}}
\setcounter{equation}{0}

One of the most important and most intensively studied problems 
in nuclear physics is deriving the forces among nucleons. 
At present, we are not able to perform direct calculations
of these starting from quantum chromodynamics (QCD). Instead of this 
the alternative approach
based on the  chiral symmetry of QCD  was pioneered by Weinberg
\cite{swnp1}\cite{weinnp} and developed by 
van Kolck et al. \cite{ubi0}\cite{ubi1}\cite{ubi2}. 
In this method one starts from
the most general Lagrangian for pions and nucleons.
Apart from the usual symmetry constraints,  spontaneously broken 
chiral symmetry 
%%%%%%%%%%%%%%%%%%%%%%%%%%%%%%%%%
is
%%%%%%%%%%%%%%%%%%%%%%%%%%%%%%%%%
taken into account. Consequently, the  couplings of pion fields,
which play the rule of Nambu-Goldstone bosons within this formalism, 
to nucleon fields are of derivative type. The most general pion--nucleon 
Lagrangian is then given by infinite series of terms with
increasing number  of derivatives and nucleon fields. Explicit
breaking of chiral symmetry, leading to the
nonvanishing but small pion mass, can also be treated systematically.

The Lagrangian  constructed in this way can then be used to derive the 
expansion of low energy S-matrix elements in powers of some low energy
scale $Q$.
To be more precise, the 4-momenta of the external pions and the 3-momenta
of the external nucleons must be no larger but of the same order as $Q$.  
It is therefore natural to treat the nucleons nonrelativistically.
More precisely, the dimensionless expansion parameter is
$Q/\Lambda_\chi$, with $\Lambda_\chi$ the scale of the spontaneous
symmetry violation. Although the value of $\Lambda_\chi$ is not precisely
known, it is related to the scale of the higher mass states in QCD,
i.e. $\Lambda_\chi$ can be chosen to be the mass of the rho
or $2\pi F_\pi \simeq 1.2$~GeV, with $F_\pi = 186\,$MeV the weak
pion decay constant.\footnote{Note that for easier comparison with the
existing literature, we use this particular normalization of $F_\pi$.} 
All other degrees of freedom such as higher mass mesons, antinucleons or
other baryons are less relevant for low--energy physics and may be 
considered as  to be integrated out of the theory. The information about 
them is then hidden in parameters of the Lagrangian.
Only the $\Delta(1232)$ resonance, which has a mass close to the one of 
the nucleon and is strongly coupled to the pion--nucleon
system, is sometimes kept explicitly. 
Once the effective Lagrangian is constructed, one can formulate the
power counting rules in order to determine the power of $Q$ for any 
given Feynmann diagram. It was shown in \cite{weinmass} that chiral symmetry
guarantees a perturbative expansion of low-energy processes because
the power of $Q$ for any given process is bounded from below. In particular,
only a  finite number of tree and loop diagrams  can contribute
to given processes at any finite order.

This scheme works fairly well for purely pionic or processes with only 
one nucleon \cite{bkmrev}\cite{ecker}. However, complications arise 
when one tries to  treat
processes with few nucleons caused by infrared divergences 
(in the static approximation for the nucleons), which destroy the 
power counting. These difficulties are in principle to be 
expected: the presence of low-energy bound states signals the failure
of perturbation theory. One way of solving this problem, first 
introduced by Weinberg \cite{swnp1}, is to use 
the time--dependent (``old-fashioned'') perturbation theory 
instead of the covariant one. The expansion of the S-matrix, 
obtained using this formalism, has the form of a Lippmann-Schwinger
equation with an effective potential, defined as a sum of all
diagrams  without pure nucleonic intermediate states. Such states
would lead in "old-fashioned" perturbation 
theory  to energy denominators, which are by a factor of $Q / m_N$ 
smaller then those from the states with pions and which destroy 
the power counting. The effective potential is free from such small 
energy denominators and can in principle
be calculated perturbativly to any given precision.
Many other equivalent schemes are known leading to same results 
like that due to Bloch and Horowitz \cite{BH} 
or the Tamm-Dancoff approximation \cite{TD}. 
The effective potential, derived using the old-fashioned time--ordered
perturbation theory, possesses one unpleasant property: it is in general
explicitly dependent on the energy of incoming nucleons and as a 
consequence of this is not Hermitean. Furthermore, the 
nucleonic wave functions are not orthonormal in this approach \cite{edga}.

One can avoid these problems
by using the method of unitary transformation. It was already
applied succesfully in cases where one has an expansion in a 
coupling constant, such as the pion--nucleon coupling, see e.g.
refs.~\cite{okuboapp}. 
Our  aim  is to show how to apply this method to the case of chiral
perturbation theory for pions and nucleons, in which the small momenta
of external particles play the role  of the expansion parameter. In 
fact, our considerations are more general since they can be applied to
any effective field theory of Goldstone bosons coupled to some massive
matter fields.

In  section~2 we give a short introduction to the Bloch--Horowitz
scheme and the method of unitary transformations as pionneered by
Okubo \cite{okubo} and by Fukuda et al. \cite{FST}. 
In section~3, we show how to apply 
this formalism to the chiral invariant effective Lagrangian 
for pions and nucleons. We use these ideas
in section~4 in order to calculate the two--nucleon force in leading 
and next-to-leading order using the
same effective Lagrangian as in \cite{ubi1} and compare our results 
with those obtained using ``old--fashioned'' perturbation theory.
We also derive the leading order three--nucleon force and shows that
it vanishes, a result first found by Weinberg \cite{swnp1}. Section~5 contains
the summary and conclusions. Various technicalities are relegated to
the appendices.

%%%%%%%%%%%%%%%%%%%%%%%%%%%%%%%%%%%%%%%%%%%%%%%%%%%%%%%%%%%%%%%%%%%%%%%
\section{Bloch-Horowitz scheme and the method of unitary transformation}
\setcounter{equation}{0}

A good introduction to both schemes is given in \cite{edga}. We briefly
repeat the main points, mostly to establish our notation and keep the
paper self--contained. Furthermore, we need the basic equations to
develop the chiral power counting scheme in the following.

We start from the Hamiltonian 
\beq 
H  = H_0 + H_I
\eeq
for pions and nucleons with $H_0 \, (H_I)$ denoting the free
(interaction) part. The whole problem for an  arbitrary number of 
interacting particles 
can be cast in the form of a Schr\"odinger equation,
\begin{equation}
\label{1}
H | \Psi \rangle = E | \Psi \rangle~.
\end{equation}
In order to solve this equation for nucleon--nucleon scattering, 
it is advantageous to 
project it onto a  subspace $| \phi \rangle$  of the 
full Fock space $| \Psi \rangle$
which contains only nucleonic states. We shall denote the remaining 
part of the Fock space by  $| \psi \rangle$: $| \Psi \rangle = 
| \phi \rangle +  | \psi \rangle$.
Let  $\eta$ and $\lambda$  be projection operators on the states 
$| \phi \rangle$ and $| \psi \rangle$
which satisfy $\eta^2 = \eta$, $\lambda^2 = \lambda$, $ \eta \lambda 
= \lambda \eta = 0$ and $\lambda + \eta = {\bf 1}$.
Eq.~(\ref{1}) can now be written in the form
\begin{equation}
\label{2}
\left( \begin{array}{cc} \eta H \eta & \eta H \lambda \\ 
\lambda H \eta & \lambda  H 
\lambda \end{array} \right) \left( \begin{array}{c} | \phi \rangle \\ 
| \psi \rangle \end{array} \right)
= E  \left( \begin{array}{c} | \phi \rangle \\ 
| \psi \rangle \end{array} \right)
\quad .
\end{equation}
In the Bloch--Horowitz scheme \cite{BH} one obtains  
from the second line of this equation an expression for the
state $| \psi \rangle$,
\begin{equation}
\label{3}
| \psi \rangle = \frac{1}{ E - \lambda H \lambda} \lambda H \eta | \Psi \rangle
\end{equation}
which can then be used to reduce the first line of eq.~(\ref{2}) to 
\begin{equation}
\label{4}
\left( H_0 + V_{{\rm eff}} ( E ) \right) | \phi \rangle  = E | \phi \rangle 
\end{equation}
with an effective potential $V_{\rm eff} (E)$ given by
\begin{equation}
\label{5}
V_{\rm eff} (E)= \eta H_I \eta + \eta H_I \lambda 
\frac{1}{E - \lambda H \lambda} \lambda H_I \eta \,\,\, .
\end{equation}
Expanding  the denominator in eq.~(\ref{5}) in powers of $H_I$ leads to
\begin{equation}
\label{6}
V_{\rm eff} (E) = \eta H_I \eta + 
\sum_{n=0}^\infty \eta H_I \lambda \frac{1}{E - H_0} 
\left( \lambda  H_I \lambda  
\frac{1}{E - H_0} \right)^n \lambda H_I \eta~.
\end{equation}
In this form it is obviously  identical with the 
result obtained from ``old-fashioned'' perturbation theory 
\cite{schweber}\cite{bonn}.

The effective Hamiltonian for nucleons can also be derived from
eq.~(\ref{2}) in another way, which 
we shall refer to as the method of unitary transformation.
Introducing new states $| \chi \rangle$ and $| \varphi \rangle $, 
which are related 
to $| \phi \rangle $ and $| \psi \rangle$ by the unitary transformation 
\begin{equation}
\label{7}
\left( \begin{array}{c} | \chi \rangle \\ | \varphi \rangle \end{array} \right)
= U^\dagger  \left( \begin{array}{c} | \phi \rangle \\ | \psi 
\rangle \end{array} \right) \quad ,
\end{equation}
one can rewrite eq.~(\ref{2}) in an equivalent form
\begin{equation}
\label{8}
U^\dagger H U \left( \begin{array}{c} 
| \chi \rangle \\ | \varphi \rangle \end{array} \right) =
E \left( \begin{array}{c} | \chi \rangle \\ 
| \varphi \rangle \end{array} \right) \quad .
\end{equation}
We can decouple the two subspaces $| \chi \rangle $ and $| \varphi \rangle $
by choosing $U$ such that the operator $U^\dagger H U$ is diagonal.
We adopt the ansatz of Okubo \cite{okubo} for $U$. It has the form
\begin{equation}
\label{9}
U = \left( \begin{array}{cc} (1 + A^\dagger A )^{- 1/2} & - 
A^\dagger ( 1 + A A^\dagger )^{- 1/2} \\
A ( 1 + A^\dagger A )^{- 1/2} & (1 + A A^\dagger )^{- 1/2} \end{array} \right)
\end{equation}
with the operator $A$ satisfying 
\begin{equation}
\label{9a}
A = \lambda A \eta~.
\end{equation}
The requirement for $U^\dagger H U$ to be diagonal leads  to
the following nonlinear equation for $A$:
\begin{equation}
\label{10}
\lambda \left( H - \left[ A, \; H \right] - A H A \right) \eta = 0~.
\end{equation}
In the case when the interaction Hamiltonian $H_I$ can be treated 
as a small perturbation, it is possible to solve Eq.~(\ref{10}) 
perturbatively to any order. For instance 
for the Hamiltonian $H$ represented by 
\begin{equation}
\label{11}
H = H_0 + \sum_{n=1}^{\infty} H_n
\end{equation}
with the index $n$ denoting the power of the coupling constant, one assumes  
the operator $A$ to be of the form
\begin{equation}
\label{12}
A = \sum_{n = 1}^\infty A_n \quad .
\end{equation}
The solution of eq.~(\ref{10}) to order $n$ is then given by 
\begin{equation}
\label{13}
A_n = \frac{1}{{\cal{E}} - H_0} \lambda \left\{ H_n + \sum_{i =
    1}^{n-1} H_i A_{n-i} - \sum_{i=1}^{n-1} A_{n-i} H_i
- \sum_{i = 1}^{n-2} \; \sum_{j =1}^{n - j - 1} A_i H_j A_{n -i-j} 
\right\} \eta
\,\,\, . 
\end{equation}  
Here, we denote the free-particle energy of the state 
$| \eta \rangle$ by $\cal{E}$.
One can see from eq.~(\ref{13}) that it is possible to find $A_n$
for every $n$ recursively, starting from $A_1$. 
The effective Hamiltonian, which operates solely in the subspace 
$| \chi \rangle$, is given by 
\begin{equation}
\label{14}
H_{\rm eff}= \eta ( 1 + A^\dagger A)^{-1/2} 
\left( H + A^\dagger H + H A + A^\dagger H A
\right) (1 + A^\dagger A)^{-1/2} \eta~,
\end{equation} 
as it follows from eq.~(\ref{8}). Expanding $(1 +
A^\dagger A )^{-1/2}$ and using eqs.~(\ref{11}),~(\ref{12}),~(\ref{13})
  one can obtain the effective 
Hamiltonian to any order in the coupling constant.

Several modifications are necessary
 by applying  the formalism described above to effective Lagrangians 
(Hamiltonians)  and in particular to chiral invariant Lagrangians.
First, the expansion in powers of a coupling constant must be 
replaced by the expansion in powers of small momenta. For doing that,
power counting rules are  
necessary. Furthermore, one expects the operator $\lambda A \eta$ to 
consist of an infinite number of terms to any order of $Q$ caused by 
infinite number of vertices in the Hamiltonian.
We now show how  these problems can be solved.

%%%%%%%%%%%%%%%%%%%%%%%%%%%%%%%%%%%%%%%%%%%%%%%%%%%%%%%%%%%%%%%%%%
\section{Application to chiral invariant Hamiltonians}
\setcounter{equation}{0}

We want first to recall the structure of the most general chiral 
invariant Hamilton density for pions and nucleons, 
\begin{equation}
\label{15}
{\cal H}={\cal H}_0 + {\cal H}_{I} \quad .
\end{equation}
We shall treat the nucleons nonrelativistically as it also 
has been done in \cite{swnp1}\cite{weinnp}\cite{ubi1}. 
The purely nucleonic part of ${\cal H}_0$
is nothing but the kinetic energy
\begin{equation}
\label{17}
{{\cal{H}}_N}_0 = - N^\dagger \frac{\vec {\nabla}^2}{2 m} N \quad , 
\end{equation}
where $m$ denotes the nucleon mass.
The mass term disappears from 
%%%%%%%%%%%%%%%%%%%%%%%%%%%%%%%%%%%
the 
%%%%%%%%%%%%%%%%%%%%%%%%%%%%%%%%%%%
Hamiltonian when one separates the
basic  time--dependence of nucleon fields (in the rest--frame) via 
\begin{equation}
\label{16}
N = {\rm e}^{ - i m t} \, \tilde N \quad .
\end{equation}
In higher orders in small momenta, which we shall not treat
here, relativistic corrections to eq.~(\ref{17}) must be taken into account 
and the free nucleonic Hamilton density looks like 
\begin{equation}
\label{18}
{{\cal{H}}_N}_0 = {{\cal{H}}_N}_0^{(2)} + {{\cal{H}}_N}_0^{(4)} 
+  {{\cal{H}}_N}_0^{(6)} + \ldots
\end{equation}
with ${{\cal{H}}_N}_0^{(2)}$ given by eq.~(\ref{17}). 
The upper indices denote the number of 
derivatives acting on $N$. These terms stem from the expansion of the
relativistic kinetic energy in powers of $1/2m$. A path integral
approach to derive  such relativistic corrections  explicitly
from the underlying relativistic theory in the one--baryon sector is
spelled out in \cite{bkkm}.
The free Hamilton density for pion fields is given by
\begin{equation}
\label{19}
{{\cal H}_\pi}_0 = \frac{1}{2} \dot{\fet{\pi}}^2 + \frac{1}{2} 
( \vec{\nabla}\fet{\pi} )^2 + 
\frac{1}{2} m_\pi^2 \fet \pi^2 \quad ,  
\end{equation}
where the '$\, \dot{}\,$' denotes the time derivative and $m_\pi$ the pion
mass. We split the interaction Hamilton density into three parts:
\begin{equation}
\label{20}
{\cal H}_{I} = {\cal H}_{\pi \pi} + {\cal H}_{N N} + {\cal H}_{\pi N}~. 
\end{equation}
The first piece describes self--interactions of pions and contains an even
number of derivatives and pion field operators and any number of 
$m_\pi^2$-factors.
The terms with two derivatives (or one $m_\pi^2$) have at least four 
pion fields.
 
The second piece in eq.~(\ref{20}) consists of four or more nucleon fields 
and any number of derivatives.
The terms of ${\cal H}_{\pi N}$ have any number of pion fields and 
either two nucleon fields and 
at least one derivative (or $m_\pi^2$) or four or more nucleon fields 
and any number of derivatives and factors of $m_\pi$.
The structure of the effective Hamiltonian is now clarified
%%%%%%%%%%%%%%%%%%%%%%%%%%%%%%%%%%%
and the explicit form we use is given in Appendix C.
%%%%%%%%%%%%%%%%%%%%%%%%%%%%%%%%%%%
Because of baryon number conservation   and the  
absence of anti--nucleons, the subspaces of the Fock space with different 
number of nucleons are automatically decoupled. 
That is why we define the state $| \phi \rangle $ as consisting of  
$N$ nucleons only. The state $| \psi \rangle$ contains then $N$ nucleons and 
at least one pion. To avoid difficulties with an  infinite number 
of vertices in the Hamiltonian, we shall split the subspace $| \psi \rangle$
into a series of orthonormal subspaces with definite numbers of pions
\begin{equation}
\label{21}
| \psi \rangle = | \psi^1 \rangle +  | \psi^2 \rangle + | \psi^3 \rangle
+ \ldots \quad .
\end{equation}
The indices denote the number of pions. 
Analogously we introduce the corresponding projectors
\begin{equation}
\label{22}
\lambda = \lambda^1 + \lambda^2 +  \lambda^3 + \ldots
\end{equation}
with the properties
\begin{equation}
\label{23}
\lambda^{i} \lambda^{j} = \delta_{i j} \lambda^{i}
\end{equation}  
Using eq.~(\ref{23}) we can rewrite eq.~(\ref{10}) in the form of 
an infinite system of equations
\begin{eqnarray}
\label{24}
&& \lambda^1 \left( H - \left[ A, H \right] - A H A \right) \eta = 0 
\nonumber \\ 
&& \lambda^2 \left( H - \left[ A, H \right] - A H A \right) \eta = 0 
\nonumber \\
&& \cdots \\
&& \lambda^i \left( H - \left[ A, H \right] - A H A \right) \eta = 0 
\nonumber \\
&& \cdots \nonumber
\end{eqnarray}
Because of  the property eq.~(\ref{9a}) of the operator $A$  each single 
equation in eqs.~(\ref{24}) can be expressed as
\begin{equation}
\label{25}
\lambda^i H_I \eta + \sum_{j=1}^\infty  \lambda^i H_I \lambda^j A \eta
+ \lambda^i H_0 \lambda^i A \eta
- \lambda^i A \eta H_I \eta - \lambda^i A \eta H_0 \eta - \sum_{j=1}^\infty
\lambda^i A \eta H_I \lambda^j A \eta = 0
\end{equation}      
To solve the system of eqs.~(\ref{24}), we make use of the usual 
philosophy of effective 
theories. We are interested only in low--energy processes. 
When the renormalization
scale 
%%%%%%%%%%%%%%%%%%%%%%%%%%%%%%%%%%
for loops
%%%%%%%%%%%%%%%%%%%%%%%%%%%%%%%%%%
is chosen to be of the same order as the low--energy scale $Q$, it is 
the only quantity with dimensionality of energy apart from the
coupling constants.
The low--energy matrix elements can then be classified by powers of 
$Q$ by use of simple dimensional analysis.
This immediately raises the following question: The projection of the
Hamiltonian in eq.~(\ref{14}) involves (in principle) an integration over all
momenta, whereas the matrix elements of  the operator $A$, which
can be found from eq.~(\ref{24}), are obtained from a low--energy
expansion. Does  this amount to a consistent procedure? The answer to this is
yes, because of the renormalization, after which all typical momenta 
in virtual processes are bounded and of the order of $Q$.

Now we make an ansatz for the operator $\lambda^{i} A \eta$, which
will be justified at the end of this section: we assume that it 
consists of 
%%%%%%%%%%%%%%%%%%%%%%%%%%%%%%
equal number of 
vertices from $H_I$ and energy denominators.
Each of these denominators 
%%%%%%%%%%%%%%%%%%%%%%%%%%%%%%
contributes a factor of $1/Q$.
It is now easy to calculate the power  $\nu_A$ of $Q$ 
of the matrix element $\langle  \psi^i | A | \phi \rangle$ using the 
usual topological identities,
\beq \label{chdim}
\langle  \psi^i | A | \phi \rangle = Q^{\nu_A} \, {\cal
  F}(Q/\Lambda_\chi, \mu /\Lambda_\chi, g_i ) \quad, 
\eeq
with $\mu$ the regularization scale (since regularization is needed
to render certain loop integrals finite) and $g_i$ collectively
denoting all coupling constants (low--energy constants).  
Each integration over virtual momenta brings three powers and each derivative 
one power of $Q$, respectively.
The pionic phase-space factors $1 / \sqrt{2 \omega}$, the 
energy denominators and $\delta$-functions give each a factor of
$Q^{-1/2}$, $Q^{-1}$ and  $Q^{-3}$, in order. 
For $L$ loops, $C$ separately connected pieces, 
%%%%%%%%%%%%%%%%%%%%%%%%%%%%%
$I_p$ internal pion lines and
%%%%%%%%%%%%%%%%%%%%%%%%%%%%% 
$V_i$ vertices of type $i$ with $d_i$ derivatives one finds
\begin{equation} 
\label{26}
\nu_A = 3 L + \sum_i V_i d_i  - I_p -\sum_i V_i - 3 ( C - 1 )
\end{equation}
%%%%%%%%%%%%%%%%%%%%%%%%%%%
According to our assumption $\sum_i V_i$ stands for the number of
energy denominators.
%%%%%%%%%%%%%%%%%%%%%%%%%%%
Here and in what follows  we do not count the overall 
$\delta$-function and the phase-space factors of the external pions.
Now we use the 
%%%%%%%%%%%%%%%%%%%%%%%%%%%
general
%%%%%%%%%%%%%%%%%%%%%%%%%%%
identities
\begin{equation}
\label{27}
L = C + I_p + I_n - \sum_i V_i \quad , 
\end{equation}
\begin{equation}
\label{28}
2 I_n + E_n = \sum_i V_i n_i \quad , 
\end{equation}
and
\beq
\label{28a}
2 I_p + E_p = \sum_i V_i p_i \quad ,
\eeq
where $E_n = 2 N$ is the number of external nucleon lines,
$E_p$ the number of external pion lines,
%%%%%%%%%%%%%%%%%%%%%%%%%%%
$I_n$ the number of internal nucleon lines
%%%%%%%%%%%%%%%%%%%%%%%%%%%
and $n_i$ $(p_i)$ the number of nucleon (pion) fields at a vertex of type $i$, 
to cast eq.~(\ref{26}) into  the form
\begin{equation}
\label{29}
\nu_A = 3 - 3 N - E_p + \sum_i V_i \, \kappa_i \quad ,
\end{equation}
with
\begin{equation}
\label{30}
\kappa_i = d_i + \frac{3}{2} n_i + p_i - 4~.
\end{equation}
%%%%%%%%%%%%%%%%%%%%%%%%%%%
Again we point out that
%%%%%%%%%%%%%%%%%%%%%%%%%%%
this result takes its form due to the ansatz that the projected
operators $\lambda^ i A \eta$  with a given number of nucleons and
pions consist of an equal number of vertices and energy denominators.  
Also, it should be mentioned that we use the number of external pion
lines to express the counting index $\nu_A$ instead of the number of
loops and of separately connected pieces as it has been done in 
\cite{swnp1}-\cite{ubi2}.  This is more natural in the
projection formalism employed here.
Let us take a closer look at vertex dimension $\kappa_i$, which is
related to the canonical field dimension. 
It is well known that all interactions can be classified with
respect to $\kappa_i$. Those with $\kappa_i < 0$ are called
relevant, with $\kappa_i = 0$ marginal and with $\kappa_i > 0$
irrelevant. The last ones are ``harmless'' and can be well treated
within low--energy effective field theories. In contrast to these,
the relevant and marginal interactions lead to complications with 
the power counting. This becomes immediately clear when one takes 
a look at eq.~(\ref{29}):
the number of possible diagrams for a given process and to a 
given order is no more bounded. Furthermore, for relevant interactions
the number $\nu_A$ is even not bounded from below.
That is why the perturbative treatment is not possible in this case.
For more details about the role of such interactions in  
effective field theories see \cite{pol}. 
Chiral symmetry does not allow any relevant or marginal
interactions in the interaction Hamiltonian $H_I$.
The minimum possible value of $\kappa_i$ for vertices in $H_I$ is one.
The corresponding vertex contains two nucleons, one pion and 
one derivative. In general, for vertices with two nucleons 
and $p_i$ pions one can easily find that $\kappa_i \geq p_i$.
The interactions with pions only have the vertex dimension
$\kappa_i \geq 2$ with $\kappa_i = 2$ corresponding to  the interaction 
with four pion fields and two derivatives.
Again, for those vertices with an even number $p_i$ of pions and without 
nucleons it follows, that $\kappa_i \geq {\rm max} ( 2, \; -2 + p_i )$.
%All such interactions have no nucleon fields, two derivatives and 
%at least four pion fields.
%All coupling constants of these terms are fixed by chiral symmetry, because the 
%only way for such vertices to enter the Hamiltonian is from  the
%expansion  of the covariant derivative in powers of pion fields,
%\begin{equation}
%\fet D_\mu = \frac{1}{1 + \fet \pi^2 / F_\pi^2} \frac{\partial_\mu 
%\fet \pi}{F_\pi}
%\end{equation}
%of the leading term $\fet D_\mu \fet D^\mu$. Here, we follow Weinberg 
%\cite{swnp1}\cite{weinnp} and use the stereographic parametrisation
%of the $SU$(2) manifold. Of course, any other parametrization leads to
%the same results for the observables.
%No interactions with $\kappa_i = -1$ are allowed by symmetries.
%However, there are many vertices with $\kappa_i =0$ like those with four 
%derivatives and only pion fields, with two nucleon fields, one derivative and any
%number of pions and with four nucleons without derivatives and pion fields.
%
Using eq.~(\ref{29}) one can determine the minimum value of 
$\nu_A$ for $\lambda^{4 k + i} A \eta$ with $k$ positive integer or zero
and $i = 0, 1, 2, 3$:
\begin{equation}
\label{31}
{\rm min} ( \nu_A ) = 3 - 3 N - 2 k \quad.
\end{equation}
All diagrams corresponding to such operators consist of
maximum possible number of pure pionic vertices with $p_i$ pion fields and  $\kappa_i = 
-2 + p_i$  
and any number of interactions with two nucleons, $p_i$ pions and $\kappa_i=p_i$ without 
intermediate pion states.
In the case $i=3$  one intermediate 
pion state may also occur.

We define the order of the operator $\lambda^{4 k + i} A \eta$
by the number $l$ ($l = 0, 1, 2, 3, \ldots$) which satisfies
\begin{equation}
\label{32}
\nu_A = 3 - 3 N - 2 k + l \quad ,
\end{equation}
and introduce the following notation: $\lambda^{4k + i} A_l \eta$.
Let us now find the power 
%%%%%%%%%%%%%%%%%%%%%%%%%%%%%
$\nu$ 
%%%%%%%%%%%%%%%%%%%%%%%%%%%%%
of $Q$ for every term in eq.~(\ref{25}). 
By $H_\kappa$ we denote a vertex from
the interaction Hamiltonian $H_I$, with the index $\kappa$ given by
eq.~(\ref{30}). Details
are relegated to appendix~A. It can be seen from 
eqs.~(\ref{33}), (\ref{35}), (\ref{36}), (\ref{39}) and (\ref{40})
that the minimum possible value of $\nu$ for eq.~(\ref{25}) is given by
\begin{equation}
\label{41}
{\rm min} (\nu) = 4 - 3N - 2k \quad .
\end{equation}

We  now show how to solve the system of eqs.~(\ref{24}) perturbatively.
For that, we define the order of eq.~(\ref{25}) to be $r \geq 0$ given by
\begin{equation}
\label{42}
r = \nu - ( 4 - 3N - 2k) \quad .
\end{equation}
In appendix~B we check what kind of operators $\lambda^{4k + i} A_l \eta$
%%%%%%%%%%%%%%%%%%%%%%%%%%%%%%%%
enter  the eq.~(\ref{25}) at each order $r$. That equation can be expressed as
%%%%%%%%%%%%%%%%%%%%%%%%%%%%%%%%
\begin{equation}
\label{43}
E ( \lambda^{4k + i} ) \lambda^{4k + i} A_{l=r} \eta = - \lambda^{4k + i} \left\{
H - \left[ A, H \right] - A H A \right\} \eta~.
\end{equation}
%%%%%%%%%%%%%%%%%%%%%%%%%%%%%%%%
The free energy is always denoted by $E$. In this case it is the pionic 
part only.
%%%%%%%%%%%%%%%%%%%%%%%%%%%%%%%%
In the curly brackets we have written in symbolic form all remaining terms 
of eq.~(\ref{25}). For $i=1, 2$ they contain only  
operators of the type $\lambda^{4 \tilde k + \tilde i} A_{l} \eta$
with
%%%%%%%%%%%%%%%%%%%%%%%%%%%%%%%%%%%%%
%%%%Hier und bis zum Ende des Kapitels ist viel korregiert worden!!!!!!!!!!!
\beq
\label{44}
 4 \tilde k + \tilde i < 4k + i \quad \mbox{for} \quad l = r \,\, , \quad
\mbox{or} \quad  l < r \quad . 
\eeq
As an example we regard 
$\lambda^{4k+i} H \lambda^{4 \tilde k + \tilde i} A_l \eta$.
For $4 \tilde k + \tilde i < 4k+i$ it follows from eqs.~(\ref{bcor1}), 
(\ref{bcor2}) that $l \leq r$. For $4 \tilde k + \tilde i = 4k+i$, 
eq.~(\ref{bcor2a}) leads to $l \leq r-2$. Finally for 
$4 \tilde k + \tilde i > 4k+i$ eqs.~(\ref{bcor2b})-(\ref{bcor4})
require $l \leq r-2$.
Eq.~(\ref{44}) is also true for $i=3$ apart from the single term
$- \lambda^{4k +3} H_1 \lambda^{4 (k+1)} A_{l=r} \eta$.
This is due to eq.~(\ref{bcor*}).
In the case  $i=0$, only the operators $\lambda^{4 \tilde k 
+ \tilde i} A_{l} \eta$, restricted by the conditions
\beq
\label{46}
 \tilde i = 0, \quad \tilde k < k \quad \mbox{for} \quad l = r  \,\, , \quad
\mbox{or} \quad  l < r \quad , 
\eeq  
can enter the right hand side of eq.~(\ref{43}).
In all cases the value of $\tilde k$ is bounded from above by the inequality
\begin{equation}
\label{48}
\tilde k \leq k + \frac{1}{6} (r-l +8)~.
\end{equation}
as can be defered from the second inequality in (\ref{bcor4}) and
the equality in (\ref{a5}).
Now it is clear how to deal with the system of equations~(\ref{24}).  
At each order $r$ one starts by solving the equations with $i=0$ and 
with increasing number $k$ to define  all operators $\lambda^{4k} A_{l=r} \eta$.
This requires knowledge of operators $\lambda^{4 \tilde k} A_{l} \eta$
of the same order $l=r$ with 
$\tilde k < k$ and a finite number of operators in
lower orders $l < r$. The next step is to solve the remaining equations
with $i =1, 2, 3$ with increasing number $4k +i$ starting from $4k+i =1$.
After that one can go to the next order $r+1$.
The number of equations to be solved in each order can be 
estimated by use of eq.~(\ref{48}) and the inequalities of appendix~B.  

To justify our ansatz about the  structure of the operator $A$, 
consider the starting equations at order $r=0$, 
given  by 
\begin{eqnarray}
\label{49}
&& E (\lambda^{4})   \lambda^{4} A_0 \eta = \lambda^{4} H_{2} \eta \quad ,\\
\label{50}
&& E (\lambda^{1}) \lambda^{1} A_0 \eta = \lambda^{1} H_1 \eta \quad ,
\end{eqnarray}
where $E (\lambda^{i})$ corresponds to the kinetic energy of the pions in 
the state $\lambda^i$. From eq.~(\ref{43}) one can see that the
unknown operator $\lambda^{4k+i} A_{l=r} \eta$ has the structure
which we assumed  at the  beginning of this section, if and only if   
the already known operators $\lambda A \eta$  entering the right hand side 
of this equation have precisely this form. 
Therefore, to proof our ansatz recursively for all operators 
$\lambda A \eta$ it is sufficient to see that it 
holds for the corresponding starting operators in eqs.~(\ref{49})
and (\ref{50}), which is obviously the case.
%%%%%%%%%%%%%%%%%%%%%%%%%%%%%%%%%%%%

%%%%%%%%%%%%%%%%%%%%%%%%%%%%%%%%%%%%%%%%%%%%%%%%%%%%%%%%%%%%%%%%%%%%%%%%%%%%
\section{Two-- and three--nucleon forces using the method of 
unitary transformation}
\setcounter{equation}{0}

We shall now practically apply the formalism described in the last section and
derive an effective Hamiltonian for nucleons in leading and 
next-to-leading orders.  As a starting point we
use the effective chiral invariant Hamiltonian for nucleons and pions, which was 
specified at the beginning of 
%%%%%%%%%%%%%%%%%%%%%%%%%%%%%
the
%%%%%%%%%%%%%%%%%%%%%%%%%%%%%
last section. Its explicit form is given
in appendix~\ref{app:H}.

\subsection{Two and many nucleon potential in the projection formalism}

First of all we look at the eq.~(\ref{14}) and define the order of 
all terms on its right hand side.
Introducing the operators 
\begin{equation}
\label{51}
L_t \equiv \eta A^\dagger_{l_t} \lambda^{4k_t +i_t} A_{{l'}_t} \eta
\quad ,
\end{equation}
%%%%%%%%%%%%%%%%%%%%%%%%%%%%
and its chiral power (the power of $Q$)
\beq 
\nu_t = 3 - 3N + \tilde \nu_t
\eeq
with 
\begin{equation}
\label{52}
\tilde \nu_t = 4k_t +2i_t +l_t +{l'}_t \quad ,
\end{equation}
evaluated via eqs.~(\ref{29}) and (\ref{vert})
one gets  contributions of the following types.
In the following $\nu$ denotes the corresponding power of $Q$. 
\begin{eqnarray}
&1. \quad & \Big[ \prod_t L_t \Big] \nonumber \\
\label{53*}
&& \quad \nu = 3 - 3N + \sum_t \tilde \nu_t \\ 
&2. \quad & \Big[ \prod_t L_t \Big] \eta H_\kappa \eta \Big[ \prod_s
L_s 
\Big] \nonumber \\
\label{53}
&& \quad \nu = 4 -3N + \kappa + \sum_t \tilde \nu_t + \sum_s \tilde \nu_s \\
&3. \quad & \Big[ \prod_t L_t \Big] \eta A^\dagger_{l_m} \lambda^{4k_m
  +i_m} H_\kappa \eta 
\Big[ \prod_s L_s \Big]
\nonumber \\
&& \mbox{and} \quad \Big[ \prod_t L_t \Big] \eta H_\kappa \lambda^{4k_m +i_m} 
A_{l_m} \eta \Big[ \prod_s L_s \Big]
\nonumber \\
\label{54}
&& \quad \nu = 4 -3N + \kappa + 2k_m +i_m +l_m +\sum_t \tilde \nu_t + \sum_s \tilde \nu_s \\
&4. \quad & \Big[ \prod_t L_t \Big] \eta A^\dagger_{l_m} 
\lambda^{4k_m +i_m} H_\kappa
\lambda^{4k_n +i_n} A_{l_n} \eta \Big[ \prod_s L_s \Big] \nonumber \\
\label{55}
&& \quad \nu = 4 -3N + \kappa +l_m +l_n +2k_m +2k_n +i_m +i_n 
+\sum_t \tilde \nu_t +\sum_s \tilde \nu_s~.
\end{eqnarray} 
%%%%%%%%%%%%%%%%%%%%%%%%%%%%%%%%%
In case of the unit operator in eq.~(\ref{14}) one has to drop the 
$\tilde \nu$'s.
%%%%%%%%%%%%%%%%%%%%%%%%%%%%%%%%%
The effective potential can be easily read off
from the effective Hamiltonian via
\beq
V_{\rm eff} = H_{\rm eff} - H_0 \quad.
\eeq
{}From now, we denote the 
contribution from 
%%%%%%%%%%%%%%%%%%%%%%%%%
the
%%%%%%%%%%%%%%%%%%%%%%%%%
free Hamiltonian $H_0$ by the pertinent
nucleonic and pionic free energies, ${\cal E}$ and $\omega$, respectively. 
Using now eqs.~(\ref{14}) and (\ref{53*}) - (\ref{55}) one can write 
the leading term 
%%%%%%%%%%%%%%%%%%%%%%%%%%%%%%%
of order $\nu = 6-3N$
%%%%%%%%%%%%%%%%%%%%%%%%%%%%%%%
of the $N$-body potential as:
\begin{equation}
\label{56}
V_{\rm  eff}^{(6-3N)} = \eta \left( H_2 + A^\dagger_0 \lambda^1 H_1 
+ H_1 \lambda^1 A_0 + A^\dagger_0
\lambda^1 \omega A_0  \right) \eta~.
\end{equation}
%%%%%%%%%%%%%%%%%%%%%%%%%
Note that in the last term only the pionic free energy $\omega$ 
contributes in lowest order.
%%%%%%%%%%%%%%%%%%%%%%%%% 
As  noted in \cite{weinnp} and \cite{ubi0}, 
there are is contribution to the potential to
order $7-3N$ because of parity invariance. 
Another way to see this is the fact that there is no
nonvanishing operator $\lambda^1 H_2 \eta$.
At next  order,  $8-3N$, one 
gets a more complicated expression for the potential:
\begin{eqnarray}
\label{57}
V_{\rm eff}^{(8-3N)} &=& \eta \bigg( H_4 + A^\dagger_2 \lambda^1 H_1 
+ A^\dagger_0 \lambda^1 H_3
+ H_3 \lambda^1 A_0 + H_1 \lambda^1 A_2 \nonumber \\ 
&& {}+ A^\dagger_0 \lambda^1 H_2 \lambda^1 A_0  
- \frac{1}{2} A^\dagger_0 \lambda^1 A_0 \eta H_2  - \frac{1}{2} H_2
\eta  A^\dagger_0 \lambda^1 A_0
\nonumber \\
&& {}+ A^\dagger_0 \lambda^1 {\cal E} A_0  
- \frac{1}{2} A^\dagger_0 \lambda^1 A_0 {\cal E} - \frac{1}{2} 
{\cal E} A^\dagger_0 \lambda^1 A_0
+ A^\dagger_2 \lambda^1 \omega A_0 + A^\dagger_0 \lambda^1 \omega A_2 \nonumber \\
&& {} + A^\dagger_0 \lambda^1 H_1 \lambda^2 A_0 + A^\dagger_0 
\lambda^2 H_1 \lambda^1 A_0 \\ 
&& {} - \frac{1}{2} A^\dagger_0 \lambda^1 A_0 \eta A^\dagger_0 \lambda^1 H_1
- \frac{1}{2} A^\dagger_0 \lambda^1 A_0 \eta H_1 \lambda^1 A_0 
- \frac{1}{2} A^\dagger_0 \lambda^1 H_1 \eta A^\dagger_0 \lambda^1 A_0
\nonumber \\
&& {} -\frac{1}{2} H_1 \lambda^1 A_0 \eta A^\dagger_0 \lambda^1 A_0
-   \frac{1}{2} A^\dagger_0 \lambda^1 A_0 \eta A^\dagger_0 \lambda^1 \omega A_0 
- \frac{1}{2} A^\dagger_0 \lambda^1 \omega A_0 \eta A^\dagger_0
\lambda^1 A_0 \nonumber \\
&& {} + A_0^\dagger \lambda^2 H_2 + H_2 \lambda^2 A_0 + A^\dagger_0 
\lambda^2  ( \omega_1  + \omega_2 ) A_0 \bigg) \eta~.
\nonumber      
\end{eqnarray} 
%Here, $\omega$ and ${\cal E}$ are the free energies of pions and
%nucleons, respectively.
%%%%%%%%%%%%%%%%%%%%%%%%%%%%
Note that the ${\cal E}$'s denote the nucleonic free energies related to the accompanying
projection operators ($\lambda$ or $\eta$).
%%%%%%%%%%%%%%%%%%%%%%%%%%%%
The operators $\lambda^1 A_0 \eta$, $\lambda^2 A_0 \eta$, $\lambda^1 A_2 \eta$
can be evaluated along the lines described in the last section.
This leads to:
\begin{eqnarray}
\label{58}
\lambda^1 A_0 \eta &=& - \frac{\lambda^1}{\omega} H_1 \eta~,  \\
\label{59}
\lambda^2 A_0 \eta &=& - \frac{\lambda^2}{\omega_1 + \omega_2} H_2
\eta -   \frac{\lambda^2}{\omega_1
+ \omega_2} H_1 \lambda^1 A_0 \eta~,  \\
\label{60}
\lambda^1 A_2 \eta  &=& - \frac{\lambda^1}{\omega} H_3 \eta  
- \frac{\lambda^1}{\omega}
H_1 \lambda^2 A_0 \eta - \frac{\lambda^1}{\omega} H_2 \lambda^1 A_0 \eta +
\frac{\lambda^1}{\omega} A_0 \eta H_1 \lambda^1 A_0 \eta  \\
&& {} + \frac{\lambda^1}{\omega} A_0 \eta H_2 \eta - {\cal E} \frac{\lambda^1}{\omega}
 A_0 \eta + \frac{\lambda^1}{\omega} A_0 \eta {\cal E}~.  \nonumber
\end{eqnarray}
Putting eqs.~(\ref{58}) - (\ref{60}) into eqs.~(\ref{56}) and
(\ref{57}) and performing  
straightforward algebraic manipulations, one obtains the potential expressed as 
\begin{eqnarray}
\label{61}
V_{\rm eff}^{(6-3N)} &=& \eta \Big( H_2 - H_1 \frac{\lambda^1}{\omega} H_1
\Big) \eta \\
\label{61a}
V_{\rm eff}^{(8-3N)} &=& \eta \Big( H_4 - H_1 \frac{\lambda^1}{\omega} H_3 -  
H_3 \frac{\lambda^1}{\omega}
H_1 \nonumber \\
&& {} + H_1  \frac{\lambda^1}{\omega} 
H_2 \frac{\lambda^1}{\omega} H_1 
+ H_1 \frac{\lambda^1}{\omega}
H_1  \frac{\lambda^2}{\omega_1 + \omega_2} H_2 + H_2 \frac{\lambda^2}
{\omega_1 + \omega_2}
H_1 \frac{\lambda^1}{\omega} H_1 
\nonumber \\
&& {} - H_2 \frac{\lambda^2}{\omega_1 + \omega_2} H_2 \\ 
&& {} - H_1 \frac{\lambda^1}{\omega} H_1 \frac{\lambda^2}{\omega_1 +
  \omega_2} H_1 
\frac{\lambda^1}{\omega} H_1 + \frac{1}{2} H_1
\frac{\lambda^1}{(\omega )^2} H_1 \eta 
H_1 \frac{\lambda^1}{\omega} H_1 + \frac{1}{2} H_1 \frac{\lambda^1}
{\omega} H_1 \eta 
H_1 \frac{\lambda^1}{(\omega )^2} H_1   \nonumber \\
&& {} 
- \frac{1}{2}
H_1 \frac{\lambda^1}{( \omega )^2} H_1 \eta H_2  - \frac{1}{2} H_2 \eta H_1  
\frac{\lambda^1}{( \omega )^2} H_1  \nonumber \\
&& {} + H_1 {\cal E} \frac{\lambda^1}{(\omega )^2} H_1 - \frac{1}{2} {\cal E} H_1 
\frac{\lambda^1}{(\omega )^2} H_1  - \frac{1}{2} H_1  
\frac{\lambda^1}{(\omega )^2} H_1
{\cal E} \Big) \eta~.  \nonumber 
\end{eqnarray}
Consider first the leading order potential, eq.~(\ref{61}). 
It consists of two terms,
the one--pion exchange $\sim H_1 (\lambda^1 / \omega) H_1$ and the 
four--nucleon contact interactions subsumed in $H_2$. This potential
obviously agrees with the one obtained in time--dependent perturbation
theory. More interesting is the first correction given in
eq.~(\ref{61a}). The diagrams corresponding to the various terms
are shown in figs.~1-4 and 6. The first line refers to the graphs
of fig.~1, the
second to 1, 2, 3 in fig.~2 and 1, 2 in fig.~6
 and the third to graph~4 of fig.~2. We should mention that
all  graphs containing vertex corrections with exactly 
one $\pi\pi NN$--vertex, which
are also contained in the second line, give no contributions,
because only odd functions of the loop momentum enter the 
corresponding integrals. The first term in the
fourth line subsumes graphs~5 to 8 of fig.~2 plus the irreducible self--energy
diagrams depicted in fig.~3. The next two terms in the fourth line
refer to graphs~9 and 10 plus the ``reducible'' self--energy diagrams of
fig.~4. Such ``reducible'' diagrams are typical for the method of
unitary transformation and do not occur in old--fashioned
time--ordered perturbation theory. They should not be confused with
truly reducible diagrams, one example being shown in fig.~5a. In that
figure, the horizontal dashed lines represent the states whose free
energy enters the pertinent energy denominators. In  old--fashioned
time--ordered perturbation theory, such reducible diagrams are
generated by iterating the potential in a Lippmann--Schwinger
equation, with the potential being defined to consist only of
truly irreducible diagrams. In contrast to the really reducible
graphs like the one in fig.~5a, the ones resulting by 
applying the projection formalism do not
contain the energy denominators corresponding to the propagation of
nucleons only. In the same notation as used for fig.~5a, one can e.g.
express diagram 9 of fig.~2 as a sum of two graphs as depicted in fig.~5b,c.
All diagrams involving four--nucleon contact interactions,
shown in fig.~6, follow from the fifth line
and, as already noted above, from the first term in the second line of eq.~(\ref{61a}). 
We remark that 
%%%%%%%%%%%%%%%%%%%%%%%%%
in the c.~m.~ system
%%%%%%%%%%%%%%%%%%%%%%%%%
the three terms in the last line add up to zero. We
have nevertheless made them explicit here since in time--ordered
perturbation theory, these terms are treated differently and lead
to the recoil correction, i.e. the explicit energy--dependence, of the
two--nucleon potential. 

\subsection{Two--nucleon potential: Expressions and discussion}

As already noted above, the leading part of two nucleon
potential is given by just one pion exchange with both vertices 
coming from eq.~(\ref{ham1}) and the nucleons treated as static particles plus  
contact interactions corresponding to eq.~(\ref{ham3}).
This result holds in both time--dependent perturbation theory and
projection formalism. For easier comparison, in this paper we
 use the same notation and definitions 
as it has been done in \cite{ubi1}. In particular, the spin
and isospin matrices $\vec \sigma$ and $\fet t$ satisfy the following
relations:
\beqa
\label{2n1}
t_i t_j &=& \frac{1}{4} \delta_{ij} + \frac{1}{2} i \epsilon_{ijk} t_k
\,\,\, ,\\  \label{2n2}
\sigma_i \sigma_j &=& \delta_{ij} + i \epsilon_{ijk} \sigma_k \,\,\, .
\eeqa
The initial (final) momentum of nucleon in c.~m.~system is denoted by
$\vec p$ ($\vec p \, '$). The transferred and average momenta are given 
by $\vec {q} = \vec {p} - \vec {p} \, '$ and $\vec k = (\vec p + \vec p \, ')/2$,
respectively.
The leading order potential is then given by
\beq
\label{2n3}
V^{(0)} = - \left( \frac{2 g_A}{F_\pi} \right)^2 
{\fet t}_1 \cdot {\fet t}_2 \frac{(\vec \sigma_1 \cdot
\vec q \, ) \, ( \vec \sigma_2 \cdot \vec q  \, )}{{\vec q \,}^2 
+ m^2_\pi} + C_S + C_T \, \vec \sigma_1 \cdot \vec \sigma_2
\eeq  
%%%%%%%%%%%%%%%%%%%%%%%%%%%
%%%Fierz Umordnung im Appendix C erklaert.
%%%%%%%%%%%%%%%%%%%%%%%%%%%

\medskip

\noindent The first corrections to this result appear at order two.
We  now enumerate all such corrections. We start with the ones
obtained using old--fashioned   
time-dependent perturbation theory \cite{ubi0}\cite{ubi1}.
First, one seemingly obtains the contributions of tree diagrams 1 and 2 in fig.~1
with one vertex coming from eq.~(\ref{ham6}) 
and still using the static approximation  for the  nucleons,
\begin{equation}
\label{2n4}
V^{(2)}_{1 \pi, \, {\rm tree}} = - \frac{2 g_A}{F_\pi^2} \fet t_1 \cdot \fet t_2
\frac{( \vec q \cdot \vec \sigma_1 ) ( \vec q \cdot \vec \sigma_2 )}
{{\vec q \,}^2 + m_\pi^2}
\left( A_1 {\vec q \,}^2 + A_2 {\vec k \,}^2 \right)~.             
\end{equation} 
The constants $A_1$ and $A_2$ are given by 
\begin{equation}
\label{2n5}
A_1 = - \left( A_1 ' + \frac{A_2 '}{2} \right)~, 
\quad A_2 = - 4 \left( A_1 ' - \frac{A_2 '}{2} \right)~.  
\end{equation}
We note that these expressions differ from the ones given in
appendix~B of \cite{ubi1}. Since $\vec k\,$ and $\vec q\,$ are
defined with a different prefactor as sum and difference of
$\vec p\,$ and $\vec p\,'$, there can not be the same prefactor
relating the $(A_1,A_2)$ to the  $(A_1',A_2')$. A few more remarks
are in order: If one decomposes ${\vec q \,}^2$ as ${\vec q \,}^2 +
m_\pi^2 - m_\pi^2$, one sees that the $A_1$--contribution can be
completely absorbed in coupling constant and contact term
redefinition. However, as discussed in appendix C, the  term 
 (\ref{ham6}) appears in 
%%%%%%%%%%%%%%%%%%%%%%%%%%%%%%%%%%%%%
the 
%%%%%%%%%%%%%%%%%%%%%%%%%%%%%%%%%%%%%
Hamiltonian with a fixed coefficient $\sim 1/m$
representing relativistic corrections and is not relevant to 
%%%%%%%%%%%%%%%%%%%%%%%%%%
the
%%%%%%%%%%%%%%%%%%%%%%%%%%
order we are
working if the nucleon mass is counted via eq.~(\ref{38}).
For this reason the contribution (\ref{2n4}) to the two--nucleon
potential can and should be completely omitted.

The  contributions from tree graphs with various contact 
interactions from eq.~(\ref{ham5}) can be expressed as
\begin{eqnarray}
\label{2n6}
V^{(2)}_{NN, \, {\rm tree}} &=& C_1 {\vec q \,}^2 
+ C_2 {\vec k \,}^2 + ( C_3 {\vec q \, }^2 + C_4 {\vec k \,}^2 ) 
( \vec \sigma_1 \cdot \vec \sigma_2 )
+ i C_5 \frac{ \vec \sigma_1 + \vec \sigma_2}{2} 
\cdot ( \vec q \times \vec k ) \nonumber \\
&& {} + C_6 ( \vec q \cdot \vec \sigma_1 ) 
( \vec q \cdot \vec \sigma_2 ) + C_7 ( \vec k \cdot \vec \sigma_1 )
( \vec k \cdot \vec \sigma_2 ) 
\end{eqnarray}
where the $C_i$'s are given by
\begin{eqnarray}
\label{2n7}
C_1 &=& C_1 ' - C_3 ' + \frac{C_2 '}{2}~, \nonumber  \\
C_2 &=& 4 C_1 ' - 4 C_3 ' - 2 C_2 '~, \nonumber \\
C_3 &=& C_9 ' + \frac{C_{12} ' }{2} - C_{14} '~, \nonumber \\
C_4 &=& 4 C_9 ' - 2 C_{12} ' + 4 C_{14} '~, \label{E.32} \\
C_5 &=& 2 C_5 ' - 4 C_4 ' - 2 C_6 '~, \nonumber \\
C_6 &=& C_7 ' + C_8 ' + \frac{C_{10} '}{ 2} + \frac{C_{11} '}{2} 
- C_{13} '~, \nonumber \\
C_7 &=& 4 C_7 ' + 4 C_8 ' - 2 C_{10} ' - 2 C_{11} ' + 4 C_{13} '~. \nonumber
\end{eqnarray}
Note again that the contributions $\sim C_{5}\, , \sim C_{11}'$ 
and $\sim C_{14}'$ and also the signs in many places
are different from what is given in \cite{ubi1}.
Further corrections arise from irreducible one--loop diagrams 1-8 in fig.~2:
\begin{eqnarray}
\label{2n8}
V^{(2)}_{2 \pi, \, {\rm 1-loop}, \, {\rm irr.}} 
&=&  \frac{4 g_A^2}{F_\pi^4} \, \int \, \frac{d^3 l}{(2 \pi )^3}  
\, \fet t_1 \cdot \fet t_2 \,
\frac{\left( {\vec l \,}^2 - {\vec q \, }^2 \right)}{\omega_+ 
\omega_- \left( \omega_+ + \omega_- \right)} 
\nonumber \\
&& {} - \frac{1}{2 F_\pi^4} \, \int  \, \frac{d^3 l}{(2 \pi )^3}  
\, \fet t_1 \cdot \fet t_2
\frac{ \left( \omega_+ - \omega_- \right)^2}{\omega_+ + \omega_-} 
\frac{1}{\omega_+ \omega_-}    \\
&& {} - \frac{g_A^4}{4 F_\pi^4} \, \int \, \frac{d^3 l}{ (2 \pi )^3} \,
\frac{1}{\omega_+^3 \omega_- } \Bigg\{ \left( \frac{3}{\omega_-} 
+ \frac{8 \fet t_1 \cdot \fet t_2}
{\omega_+ + \omega_-} \right) \left( {\vec l \,}^2 
- {\vec q \,}^2 \right)^2 \nonumber \\
&& {} + 4 \left( \frac{3}{\omega_+ + \omega_-} 
+ \frac{8 \fet t_1 \cdot \fet t_2}{\omega_-} \right)
( \vec \sigma_2 \cdot [ \vec q \times \vec l \, ] ) 
( \vec \sigma_1 \cdot [ \vec q \times \vec l \, ] ) \Bigg\} \; , \nonumber
\end{eqnarray}
where
\begin{equation}
\label{2n8a}
\omega_\pm \equiv \sqrt{ ( \vec q \pm \vec l \, )^2  + 4 m_\pi^2}~.
\end{equation}
We have also calculated the contributions from various  irreducible
one--loop diagrams 1-8 in fig.~3 and 1-2 in fig.~6, which
involve self--energy insertions and vertex corrections.
These have not been considered in \cite{ubi0}\cite{ubi1}.
The corresponding contributions are given by
\beqa
\label{2n9}
V^{(2)}_{1 \pi,\, {\rm 1-loop}, \, {\rm irr.}} &=&  
 \frac{4 g_A^4}{F_\pi^4} \, \int \, \frac{d^3 l}{(2 \pi )^3} \,
\frac{\fet t_1 \cdot \fet t_2}{\omega_l^2 \omega_q^2} \, 
\left\{ \frac{\vec l \cdot \vec q}{\omega_l} 
\Big( ( \vec \sigma_1
\cdot \vec l \, ) \, ( \vec \sigma_2 \cdot \vec q \, ) 
+ ( \vec \sigma_1 \cdot \vec q \, ) 
\, ( \vec \sigma_2 \cdot \vec l \, ) 
\Big) \right. \nonumber \\
&&   -  \left. \left( \frac{1}{\omega_l} 
+ \frac{3}{\omega_q} \right) {\vec l \,}^2 \, ( \vec \sigma_1 \cdot \vec q \, ) \,
( \vec \sigma_2 \cdot \vec q \, ) \right\} \;, 
\eeqa
where 
\beq
\omega_q = \sqrt{ {\vec q \,}^2 + m_\pi^2} \,\, , \quad
\omega_l = \sqrt{ {\vec l \,}^2 + m_\pi^2} \,\, ,
\eeq
and by
\beqa
\label{2n10}
V^{(2)}_{NN, \, {\rm 1-loop},  \, {\rm irr.}} 
&=& \frac{g_A^2}{F_\pi^2} C_S
\, \int \, \frac{d^3 l}{(2 \pi )^3}
\, \frac{1}{\omega_l^3} \Big\{ 3 {\vec l \,}^2  + 4 ( \fet t_1 \cdot \fet t_2 ) (\vec l \cdot \vec \sigma_1 )
( \vec l \cdot \vec \sigma_2 ) \Big\} \\
&&  + \frac{g_A^2}{F_\pi^2} C_T \, \int \, \frac{d^3 l}{(2 \pi )^3} \,
\frac{1}{\omega^3_l}
\bigg\{ 4 {\vec l \,}^2 \fet t_1 \cdot \fet t_2 + 
2 ( \vec l \cdot \vec \sigma_1 ) ( \vec l \cdot \vec \sigma_2 )
(3 - 2 \fet t_1 \cdot \fet t_2 ) \nonumber \\ 
&&  - {\vec l \,}^2 ( \vec \sigma_1 \cdot \vec \sigma_2 ) 
(3 - 4 \fet t_1 \cdot \fet t_2 )
\bigg\}~. \nonumber
\eeqa
We remark an unpleasant feature of 
%%%%%%%%%%%%%%%%%%%%%%%%%%%%%%%%%
$V^{(2)}_{1 \pi,\, {\rm
    1-loop},  \, {\rm irr.}}$. 
%%%%%%%%%%%%%%%%%%%%%%%%%%%%%%%%%
With  $t =- {\vec q\,}^2$ one can rewrite the term
proportional to $1/\omega_q^3$ as $(m_\pi^2-t)^{-3/2}$. This function
has a cut  starting at $t = m_\pi^2$. Physically, this does not
make sense since all cuts should be produced by multi--particle
intermediate states. We come back to this later on.

Finally, the corrections arising by explicitly keeping the nucleon kinetic energy 
in one--pion exchange tree diagrams with both vertices from leading 
order Lagrangian represents the 
energy--dependent part of 2N--potential given by
\begin{equation}
\label{2n11}
V^{(2)}_E = \frac{4 g_A^2}{F_\pi^2} \fet t_1 \cdot \fet t_2 ( \vec \sigma_1 \cdot \vec q \, ) 
( \vec \sigma_2 \cdot \vec q \,)
\frac{E - \frac{1}{m} \left( {\vec k \,}^2 + \frac{1}{4} {\vec q \,}^2 \right)}
{\left( {\vec q \,}^2 + m_\pi^2 \right)^{3 / 2}}~,
\end{equation}
where $E$ is an initial energy of the two nucleons (full energy). It
should be stressed that on--shell, this contribution vanishes. This
on--shell center--of--mass kinematics is often used as an approximation
for calculations of NN scattering or  few--nucleon forces. 
However, if one iterates this potential in
e.g. a Lippmann--Schwinger equation, the energy $E$ can no longer be
simply related to the momenta ${\vec k \,}$ and ${\vec q \,}$. A
similar comment applies to using this recoil correction in the
calculation of few--nucleon forces (for $N\geq 3$). 

\medskip

\noindent Let us now look at the corrections arising in the framework of 
the projection formalism.
The corrections from tree diagrams with one vertex taken from
next--to--leading order Lagrangian do not change and are given again 
by eq.~(\ref{2n6}). Remember that we do not consider  the Hamiltonian 
${\cal H}_3$ given by eq.~(\ref{ham6}) for reasons given above and in 
appendix~C. Consequently, there is no contribution (\ref{2n4}) to the 
potential.
Apart from the corrections from irreducible one--loop graphs 1-8 in
fig.~2 given by eq.~(\ref{2n8}) one obtains a contributions from 
reducible diagrams 9 and 10, which can be expressed by
\beqa
\label{2n12}
V^{(2)}_{2 \pi, \, {\rm 1-loop}, \, {\rm red.}} 
&=&  \frac{2 g_A^4}{F_\pi^4} \, \int \, \frac{d^3 l}{(2 \pi )^3}
\left( \frac{3}{4} - 2 \fet t_1 \cdot \fet t_2 \right) \frac{\omega_+ + \omega_-}
{\omega_+^3 \omega_-^3} \\ 
&& \times \Bigg\{ \frac{1}{4} \left( {\vec l \,}^2 - {\vec q \,}^2 \right)^2
- ( \vec \sigma_2 \cdot [ \vec q \times \vec l \, ] ) ( \vec \sigma_1 
\cdot [ \vec q \times \vec l \, ] ) 
\Bigg\} \, \, .  \nonumber
\eeqa
Summing up the eqs.~(\ref{2n8}) and (\ref{2n12}) we obtain the 
two--pion exchange contributions to the potential within the projection 
formalism:
\begin{eqnarray}
\label{2n12a}
V^{(2)}_{2 \pi, \, {\rm 1-loop}} 
&=&  \frac{4 g_A^2}{F_\pi^4} \, \int \, \frac{d^3 l}{(2 \pi )^3}  
\, \fet t_1 \cdot \fet t_2 \,
\frac{\left( {\vec l \,}^2 - {\vec q \, }^2 \right)}{\omega_+ 
\omega_- \left( \omega_+ + \omega_- \right)} 
\nonumber \\
&& {} - \frac{1}{2 F_\pi^4} \, \int  \, \frac{d^3 l}{(2 \pi )^3}  
\, \fet t_1 \cdot \fet t_2
\frac{ \left( \omega_+ - \omega_- \right)^2}{\omega_+ + \omega_-} 
\frac{1}{\omega_+ \omega_-}    \\
&& {} - \frac{g_A^4}{ F_\pi^4} \, \int \, \frac{d^3 l}{ (2 \pi )^3} \,
\frac{\omega_+^2 + \omega_+ \omega_- + \omega_-^2}
{\omega_+^3 \omega_-^3 (\omega_+ + \omega_- )} \Bigg\{  
2 \fet t_1 \cdot \fet t_2
\left( {\vec l \,}^2 
- {\vec q \,}^2 \right)^2 \nonumber \\
&& {} + 3 ( \vec \sigma_2 \cdot [ \vec q \times \vec l \, ] ) 
( \vec \sigma_1 \cdot [ \vec q \times \vec l \, ] ) \Bigg\} \; . \nonumber
\end{eqnarray}
As was first noted in ref.~\cite{norb} using a different formalism, 
the isoscalar spin independent central  and the isovector spin dependent 
parts of the two--nucleon potential corresponding to the  
two--pion exchange adds up to zero.
It is comforting that we find the same
result.
Note, that this is different from the energy--dependent potential derived 
using the time--ordered perturbation theory.

The corrections arising from one--loop reducible diagrams 
1-4 in fig.~4 and 3-6 in fig.~6,
which involve the nucleon self--energy and loop corrections to the
four--fermion interactions, are given by
\beq
\label{2n13}
V^{(2)}_{1 \pi,\, {\rm 1-loop}, \, {\rm red.}} =   \frac{12 g_A^4}{F_\pi^4} 
\, \int \, \frac{d^3 l}{(2 \pi )^3} \,
\fet t_1 \cdot \fet t_2 \frac{\omega_l + \omega_q}{\omega_l^3 \omega_q^3} \,
{\vec l \,}^2 \, ( \vec \sigma_1 \cdot \vec q \, ) \,
( \vec \sigma_2 \cdot \vec q \, ) 
\eeq
and by 
\beqa
\label{2n14}
V^{(2)}_{NN, \, {\rm 1-loop},  \, {\rm red.}} &=& 
- \frac{g_A^2}{F_\pi^2} C_S \, \int \, \frac{d^3 l}{(2 \pi )^3}
\, \frac{1}{\omega_l^3} \Big\{ 3 {\vec l \,}^2  + 4 ( \fet t_1 \cdot \fet t_2 ) (\vec l \cdot \vec \sigma_1 )
( \vec l \cdot \vec \sigma_2 ) \Big\} \\
&&  - \frac{g_A^2}{F_\pi^2} C_T \, \int \, \frac{d^3 l}{(2 \pi )^3} \,
\frac{1}{\omega^3_l}
\bigg\{ 4 {\vec l \,}^2 \fet t_1 \cdot \fet t_2 + 4 ( \fet t_1 \cdot \fet t_2 )
( \vec l \cdot \vec \sigma_1 ) ( \vec l \cdot \vec \sigma_2 )
\nonumber \\ 
&&  + {\vec l \,}^2 ( \vec \sigma_1 \cdot \vec \sigma_2 ) 
(3 - 4 \fet t_1 \cdot \fet t_2 )
\bigg\}~. \nonumber
\eeqa
Note that 
%%%%%%%%%%%%%%%%%%%%%%%%%%%%
$V^{(2)}_{1 \pi,\, {\rm 1-loop}, \, {\rm red.}}$ 
%%%%%%%%%%%%%%%%%%%%%%%%%%%%
again contains a term $\sim
1 / \omega_q^3$. Combining eqs.~(\ref{2n9}) and (\ref{2n13}), this unphysical
contribution vanishes.
Summing up the corrections eqs.~(\ref{2n9}), (\ref{2n10}), 
corresponding to irreducible graphs,
%%%%%%%%%%%%%%%%%%%%%%%%%%%%%%%%%%%%%%%%%%
which are again the same as in the old--fashioned time--dependent perturbation theory
%%%%%%%%%%%%%%%%%%%%%%%%%%%%%%%%%%%%%%%%%%
and those (\ref{2n13}), (\ref{2n14}), 
corresponding to reducible diagrams, one gets the 
complete result, representing the one--loop contributions, 
which involve self--energy insertions 
and vertex corrections within projection formalism:
\beqa
\label{2n15}
V^{(2)}_{1 \pi,\, {\rm 1-loop} } &=&  
\frac{4 g_A^4}{F_\pi^4} \, \int \, \frac{d^3 l}{(2 \pi )^3} \,
\frac{\fet t_1 \cdot \fet t_2}{\omega_l^3 \omega_q^2} \, 
\Bigg\{ ( \vec l \cdot \vec q \, )  
\Big( ( \vec \sigma_1
\cdot \vec l \, ) \, ( \vec \sigma_2 \cdot \vec q \, ) 
+ ( \vec \sigma_1 \cdot \vec q \, ) 
\, ( \vec \sigma_2 \cdot \vec l \, ) 
\Big)  \nonumber \\
&& \qquad \qquad  +   2 {\vec l \,}^2 \, ( \vec \sigma_1 \cdot \vec q \, ) \,
( \vec \sigma_2 \cdot \vec q \, ) \Bigg\} \,\,\, , \\
\label{2n16}
V^{(2)}_{NN, \, {\rm 1-loop}} &=&  \frac{8 g_A^2}{F_\pi^2} C_T \, 
\int \, \frac{d^3 l}{(2 \pi )^3} \, {1 \over \omega_l^3}
\, \left( \frac{3}{4} - \fet t_1 \cdot \fet t_2 \right) \bigg(
( \vec l \cdot \vec \sigma_1 ) ( \vec l \cdot \vec \sigma_2 ) -
{\vec l \,}^2 ( \vec \sigma_1 \cdot \vec \sigma_2 ) 
\bigg)~. 
\eeqa
%%%%%%%%%%%%%%%%%%%%%%%%%%%%%%%%%%%%%%%%%%
The complete potential of second order  in the projection formalism is 
therefore given by the expressions (\ref{2n6}), (\ref{2n12a}), (\ref{2n15})
and (\ref{2n16}).
%%%%%%%%%%%%%%%%%%%%%%%%%%%%%%%%%%%%%%%%%%

We remark that one can perform the $l$--integrations in the last
two equations and one finds that the contribution from eq.~(\ref{2n15})
can be entirely absorbed in coupling constant and wave function
renormalization 
%%%%%%%%%%%%%%%%%%%%%%%%%
whereas 
%%%%%%%%%%%%%%%%%%%%%%%%%
eq.~(\ref{2n16}) merely amounts to the
renormalization of the lowest order four--fermion contact terms.  
Finally, the crucial difference between the two formalisms consists
in the treatment of the energy--dependent term eq.~(\ref{2n11}).
As already noted above, it does not appear in the method of unitary
transformation. We note that many of the results derived here have
already been found in \cite{norb}, where one-- and two--pion exchange graphs
were calculated by means of Feynman diagrams and using dimensional
regularization.  That potential can and has been applied for
scattering processes but can not be used to solve the bound--state
problem.

\subsection{The three--nucleon potential}

Let us consider the leading part of 
%%%%%%%%%%%%%%%%%%%%%%%
the 
%%%%%%%%%%%%%%%%%%%%%%%
three--nucleon (3N)--potential.
For this we take a closer look at eq.~(\ref{61a}).
The diagram 1 in  fig.~7 represents all terms in the second
line of this equation, which involve the nonlinear $\pi \pi NN$
coupling. These contributions are identical in both time--dependent 
perturbation theory and in projection formalism and, as it was noted 
in ref.~\cite{swnp1},
sum up to zero, when all time--orderings are taken into account.
For three--nucleon forces mediated by two--pion exchange with only linear 
$\pi NN$ couplings one 
gets different results
by using the two approaches.
The irreducible graphs 1--4 in fig.~8 represent the result of old--fashioned 
time--dependent perturbation theory.
Apart from them, the reducible graphs~5 and 6 in fig.~8 occur using
our approach. An exact cancellation between the contributions 
from those reducible and irreducible diagrams was recently pointed out 
in \cite{edga} for the case of an expansion in the pion--nucleon
coupling constant and adopting the nonrelativistic approximation for
nucleons. At leading order, these two very different expansions lead
to the same set of diagrams. We  remark, however, that only for the
method employed here one has a consistent scheme of treating the
next-to-leading order corrections. Note furthermore that fig.~8
differs from the analogous one given in \cite{edga} by the topology
of diagram~6. The graphs e and g shown in fig.~3 of \cite{edga} lead
in general  to different prefactors multiplying the energy denominators. 
To understand why these  cancellations appear, let us take a closer look
at the terms in the fourth line of eq.~(\ref{61a}).
We just rewrite them schematically by pulling out the common factor $M$,
representing the spin, isospin and momentum structure, which is obviously
the same for all of these terms. 
For the irreducible diagrams we then have 
\beq
- \left[ \frac{2}{\omega_1 (\omega_1 + \omega_2) \omega_2} +   
\frac{1}{\omega_1^2 ( \omega_1 + \omega_2 )} + \frac{1}{( \omega_1 
+ \omega_2) \omega_2^2}
\right] M = - \frac{\omega_1 + \omega_2}{\omega_1^2 \omega_2^2} M~.
\eeq
The contribution from the reducible diagrams can be expressed as
\beq
\left[ \frac{1}{\omega_1^2 \omega_2} + \frac{1}{\omega_1 \omega_2^2} \right] M =
\frac{\omega_1 + \omega_2}{\omega_1^2 \omega_2^2}  M~.
\eeq
The cancellation is now evident.
The same sort of cancellation can be observed for diagrams
2, 3, 4 in fig.~7 and for 1--loop graphs 1--6 in fig.~6 with the
contact vertex $(1 / 2) C_S (N^\dagger N ) ( N^\dagger N)$ 
%and graphs 8,9,10 of fig.2 
contributing to the
two--nucleon potential, which correspond to the first term 
in the second line  and the terms in the fifth line of eq.~(\ref{61a}).
We conclude that there is no three--nucleon force at the order $\nu = 1$.
This agrees with the finding of Weinberg using time-ordered
perturbation theory~\cite{swnp1}. In that case, the mechanism is
different. Although the leading three--nucleon force corresponding to
diagrams 2 in fig.~7 and 1-4 in fig.~8 does not vanish in that case,
it was noted in \cite{swnp1} that these contributions cancel exactly
with the energy--dependent part of the two--nucleon potential when the
latter is iterated in the Lippmann--Schwinger equation. That is the
reason why one can describe systems of three or more nucleons using
the energy--independent part of the 2N--potential and without explicit
three-- and many--nucleon forces to the order in
the chiral expansion also considered here.  However, such type of
cancellations does not help to remove the problems in the two--nucleon
sector due to the explicit energy--dependence as noted before
(i.e. that the wavefunctions are only orthonormal to the order one
is working). This is the main difference to the projection approach
we are using.

%%%%%%%%%%%%%%%%%%%%%%%%%%%%%%%%%%%%%%%%%%%%%%%%%%%%%%%%%%%%%%%%%%%%%%%%
\section{Summary and conclusions}
\setcounter{equation}{0}

In this paper, we have presented a novel approach to the problem
of deriving the forces between few (two, three, $\ldots$) nucleons
from effective chiral Lagrangians. For that, we use the well--known
projection formalism dating back to 
Okubo, Fukuda, Sawada and Taketani~\cite{okubo}\cite{FST}.
It allows to uniquely separate the effects from the operators and the
wave functions. For the case at hand, we first had to modify the power
counting rules since in the projection formalism one decomposes the
Hilbert space of nucleons and pions into subspaces  with definite
nucleon and pion numbers. While in old--fashioned time--ordered
perturbation theory the resulting wave functions are only orthonormal 
to a certain order in the chiral expansion, this problem does not
occur in the projection formalism. Furthermore, in the previous
calculations based on time--ordered perturbation theory, the two--nucleon
potentials turn out to be energy--dependent, which is a severe
complication for applying these in systems with three or more
nucleons (although to leading order, these recoil corrections are
cancelled by certain N--body interactions). 
Here, we have given the explicit expressions of the
two--nucleon potential to two orders beyond the lowest order tree graphs
(comprising the leading one--pion exchange and four--nucleon
contact terms). In particular, we have also calculated self--energy
type corrections and one--loop corrections to these four--fermion
operators not considered before. The most salient features of the so 
constructed two--nucleon potential is its {\it energy}--{\it
independence} and the orthonormality of the corresponding wave functions.
Furthermore, we have considered
the leading contributions to the three--nucleon potential. As in
time--ordered perturbation theory, we find that these sum up to
zero. However, the mechanism for the cancellation between some of
the graphs  is distinctively different in the projection formalism since it
is not related to an intricate cancellation of terms generated by
the iteration of the energy--dependent two--nucleon potential in
old--fashioned time--ordered perturbation theory. Rather, in the
approach here, these cancellations can be traced back to the
appearance of ``reducible'' graphs whose precise meaning is explained
in section~4.1. These diagrams are in fact responsible for the
orthonormality of the wave functions and are thus sometimes called ``wave
function re--orthonormalization'' graphs.

Clearly, the results presented here constitute only the first step in
a bigger program. We are presently working on a regularization
scheme, which allows us to consistently treat the potential {\it and}
the Lippmann--Schwinger equation to generate the bound
state (for a recent discussion with many references to other
approaches, see e.g. Beane et al.~\cite{bcp}). In the pioneering work
by van Kolck et al.~\cite{ubi1}, this was not done but rather a
Gaussian cut--off for the momenta $\vec{q}$ and $\vec{k}$  was introduced
irrespective of the power counting. Furthermore, 
the $\Delta(1232)$ degrees of freedom have to be 
incorporated in a manner analogous to what was done for the single
nucleon case by Hemmert et al.~\cite{hhk} and finally, numerical
results have to be obtained for the two-- and three--nucleon systems.
%In particular, the low--energy constants related to the pion--nucleon
%sector (like e.g. $A_1$, $A_2$) 
%should be fixed from the available
%scattering data.
We believe that within the framework outlined here, one can find out
to what extent chiral symmetry plays a role for the nuclear forces
(the role of one-- and two--pion exchanges for the higher NN partial
waves was recently elucidated in \cite{robi}\cite{norb}). Furthermore,
this approach can also be used to calculate meson--exchange currents
in a manner which is completely consistent with the so--constructed
nucleon potential.

\section*{Acknowledgements}

We thank Norbert Kaiser for some very useful comments and Jamie Eden
for some helpful discussions.

\bigskip

%%%%%%%%%%%%%%%%%%%%%%%%%%%%%%%%%%%%%%%%%%%%%%%%%%%%%%%
\appendix
\def\theequation{\Alph{section}.\arabic{equation}}
\setcounter{equation}{0}
\section{Chiral counting for the projected equations}

In this appendix, we work out the chiral power of the different
terms appearing in eq.~(\ref{25}). Any one of these can be 
characterized by a counting index $\nu$, i.e. is proportional
to $Q^\nu$. This index should not be confused with the one
giving the chiral dimension of the matrix--elements $\langle
\Psi^i | A | \phi\rangle$ discussed in the text.
The free part of the Hamiltonian $H_0$ gives the energy, which
we denote throughout by $E$. Remember that $H_\kappa$ denotes
a term from the interaction Hamiltionian $H_I$ with the
vertex dimension $\kappa$.

%%%%%%%%%%%%%%%%%%%%
One 
%%%%%%%%%%%%%%%%%%%%
can determine the chiral power $\nu$ for all terms entering 
eq.~(\ref{25})  by using of the corresponding result (\ref{29}) for $\nu_A$,
which,  however,  should be modified to take into account the correct 
number of energy denominators
in each  case,
%%%%%%%%%%%%%%%%%%%%%%%%%%%%%%%%%
because of the presence of the operators $H_0$ and $H_I$.
%%%%%%%%%%%%%%%%%%%%%%%%%%%%%%%%%
 Furthermore, one 
%%%%%%%%%%%%%%%
uses
%%%%%%%%%%%%%%%
 eqs.~(\ref{29}), (\ref{32})
to express the sum $\sum_i V_i \kappa_i^A$ over all vertices 
entering the operator $\lambda^{4k +i} A_l \eta$ in terms of $k$, $i$, and $l$
as follows:
\beq
\label{vert}
\sum_i V_i \kappa_i^A = 2k + i + l \quad .
\eeq
For example, for $\lambda^{4k+i} H_\kappa \lambda^{4q+j} A_l \eta$ one has
\beq
\label{example}
\nu = 3 - 3N +1 - (4k+i) + \kappa + \sum_i V_i \kappa_i^A 
= 4 - 3N - 4k -i +2q +j  +l +\kappa
\eeq
where $(4k+i)$ corresponds to the number of external pion lines and 
we have added $+1$ because there is  one energy denominator less than it 
was counted in eq.~(\ref{29}).  
Proceeding analogously one can derive a corresponding chiral power 
for all other terms in eq.~(\ref{25}), which we shall use in what follows.
 
Now we take a look at the possible vertices in $H_I$ 
%%%%%%%%%%%%%%%%%%%%%%%%%%%%%
which are explained in section 3
%%%%%%%%%%%%%%%%%%%%%%%%%%%%%
with $p$ pions in order to
obtain few useful inequalities for determining the
corresponding vertex dimension $\kappa$:
\begin{eqnarray}
\label{kappa}
&& \kappa \geq 2, \quad \mbox{for} \; p=0 \,\,\, ,\label{kappa1} \\
&& \kappa \geq p, \quad \mbox{for} \; p=1, 2, 3 \,\,\, .\label{kappa2} 
\end{eqnarray}
%%%%%%%%%%%%%%%%%%%%%%%%%%%%%%%%%%%%%%%%%%%%%%%%%%%%%%%%%%%%%%%
These follow from (\ref{30}). For $p=0$
$n$ must be larger equal 4 and for $p=1,2,3$ $n$ must be 
greater equal 2 and there must be at least one derivative.
Further 
\begin{eqnarray}
&& \kappa \geq 1 \,\,\, ,\label{kappa3} \\
&& \kappa \geq -2 + p\,\,\, . \label{kappa4}
\end{eqnarray}
Whereas (\ref{kappa3}) is generally valid, (\ref{kappa4})
follows from simple inspection of eq.~(\ref{30}):
for $n=0$ there are at least two derivatives and for
$n \geq 2$ it is immediately apparent.
%%%%%%%%%%%%%%%%%%%%%%%%%%%%%%%%%%%%%%%%%%%%%%%%%%%%%%%%%%%%%%%%

We find the following terms with the corresponding
indices $\nu$:
\begin{itemize}
\item $\lambda^{4k + i} H_\kappa \eta$
\begin{equation}
\label{33}
\nu = 4 - 3N - 4k -i + \kappa~.
\end{equation}
%%%%%%%%%%%%%%%%%%%%%%
This follows from eq.~(\ref{29}) noting that there is one energy denominator less.
%%%%%%%%%%%%%%%%%%%%%%
It follows from (\ref{kappa1}), (\ref{kappa2}), (\ref{kappa4}), that 
\begin{description}
\item[] $\nu \geq 4 - 3N$, when $k = 0$
\item[] $\nu \geq 2 - 3N$, when $k > 0$
\end{description}
\item $\lambda^{4k + i} H_\kappa
 \lambda^{4q + j} A_l \eta$
\begin{equation}
\label{34}
\nu = 4 - 3 N -4k -i + 2q +j +\kappa + l~.
\end{equation}
The following cases are to be considered:
\begin{enumerate}
\item
$q \leq k -1$

%%%%%%%%%%%%%%%%%%%%%%%%%%%%
In this case the number of pions $p \geq 4k+i-4q-j$ at the vertex.
Therefore
%%%%%%%%%%%%%%%%%%%%%%%%%%%%
the inequality (\ref{kappa4}) takes the form
\beq
\label{34a}
\kappa \geq -2 +4k +i -4q -j \quad .
\eeq
It then follows from eq.~(\ref{34}), that
\beq
\label{35}
\nu \geq  4 - 3N - 2k \quad .
\eeq
\item 
$q=k$
 
%%%%%%%%%%%%%%%%%%
Then 
\beq
\label{34a2}
p \geq | j-i | \quad .
\eeq
We apply eq.~(\ref{kappa2}) to this case and using 
%%%%%%%%%%%%%%%%%%
the (obvious) inequality $j-i \geq - | j-i |$ one 
can justify eq.~(\ref{35}).
%\begin{enumerate}
%\item 
%$\kappa \geq 0 \quad \Rightarrow  \quad y \geq 0$
%\item 
%$\kappa \geq -2 \quad \Rightarrow  \quad y \geq 1$
%\end{enumerate}
\item 
$ q \geq k + 1$

One can easily prove 
%%%%%%%%%%%%%%%%%%
eq.~(\ref{35}) 
%%%%%%%%%%%%%%%%%%
for this case using
eq.~(\ref{kappa3}) and by noting, that $j-i \geq -3$.
%\begin{enumerate}
%\item
%$\kappa \geq 0 \quad \Rightarrow  \quad j - i \geq -3$
%\item
%$\kappa = -2 \quad \Rightarrow  \quad j - i \geq -2$
%\end{enumerate}
\end{enumerate}
%Using (\ref{34}) we conclude that
%\begin{equation}
%\label{35}
%\nu \geq 4 -3N -2k
%\end{equation}
\item $E ( \lambda^{4 k + i} )  
\lambda^{4k + i} A_l \eta$
\begin{equation}
\label{36}
\nu \geq 4 - 3N - 2k + l \geq 4 -3N -2k \quad .
\end{equation}
%%%%%%%%%%%%%%%%%%%%%%%%%%%%%%%%%%
This follows immediately from eq.~(\ref{32})
%%%%%%%%%%%%%%%%%%%%%%%%%%%%%%%%%%
Here the first inequality becomes the exact equality when only the pion 
kinetic energy is taken into account in $E ( \lambda^{4 k + i} )$.
\item $\lambda^{4k + i} A_l \eta H_\kappa \eta$
\begin{equation}
\label{37}
\nu = 4 - 3N - 2k + l + \kappa \quad .
\end{equation}
There is no nonvanishing operator $\eta H_\kappa \eta$ with $\kappa < 0$.
That is why the inequality (\ref{35}) is again valid.
\item $\lambda^{4k + i} A_l \eta E (\eta)$

We shall count the nucleon mass in the same way
as it has been done by Weinberg \cite{weinnp}:
\begin{equation}
\label{38}
\frac{Q}{m} \sim \frac{Q^2}{\Lambda_{\chi}^2} \quad .
\end{equation}
The nucleon kinetic energy must then be counted as $Q^3$ and we get the following 
result:
\begin{equation}
\label{39}
\nu \geq 6 - 3N - 2k + l \geq 6 - 3N - 2k \quad . 
\end{equation}
\item $\lambda^{4k + i} A_{l_1} \eta H_\kappa 
\lambda^{4q + j} A_{l_2} \eta$ 
\begin{equation}
\label{40}
\nu = 4 - 3N  - 2k + 2q +j + l_1 + l_2 + \kappa \geq 6 - 3N -2k \quad .
\end{equation}
\end{itemize}

%%%%%%%%%%%%%%%%%%%%%%%%%%%%%%%%%%%%%%%%%%%%%%%%%%%%%%%%%%%%%%%%%%%%%%%%%%%%

\setcounter{equation}{0}
%%%%%%%%%%%%%%%%%%%%%%
\section{Operators  at order $r$ in Eq.~(3.11)}
%%%%%%%%%%%%%%%%%%%%%%
\label{cmss}

In this appendix, we work out in detail which operators 
actually appear in eq.~(\ref{25})
%%%%%%%%%%%%%%%%%%%%%%%%%%%%%%%%%%%%%
at order $r$, which is defined in eq.~(\ref{42}). 
%%%%%%%%%%%%%%%%%%%%%%%%%%%%%%%%%%%%%
We find
\begin{itemize}
\item $\lambda^{4k + i} H_\kappa \lambda^{4q + j} A_l \eta$

One 
%%%%%%%%%%%%%
obtaines
%%%%%%%%%%%%%
 from (\ref{34}) and (\ref{42}) the following identity:
\begin{equation}
\label{a1}
l = r + 2 ( k-q ) +i -j - \kappa \quad .
\end{equation}
Let us first consider the case  $i \neq 0$:
\begin{enumerate} 
\item $q \leq k-1$

Using eqs.~(\ref{34a}), (\ref{a1}) one gets
\beq
\label{bcor1}
l \leq r - 2k + 2q + 2 \leq r~.
\eeq
\item $q=k$
\begin{enumerate}
\item
$j < i$

It follows from eqs.~(\ref{34a2}), (\ref{a1}), that
\beq
\label{bcor2}
l \leq r - |j - i| + i - j \leq r~.
\eeq 
\item
$j=i$

%%%%%%%%%%%%%%%%%%%%
In this case eq.~(\ref{a1}) 
takes the form
\beq
\label{bcor2a}
l = r-\kappa \leq r-2~.
\eeq
due to the inequalities (\ref{kappa1}), (\ref{kappa2}) and (\ref{kappa4}).
%%%%%%%%%%%%%%%%%%% 
\item
$j \geq i+1$

%%%%%%%%%%%%%%%%%%%%%%%%%%%%
The first inequality of eq.~(\ref{bcor2}) 
%%%%%%%%%%%%%%%%%%%%%%%%%%%%
leads to 
\beq
\label{bcor2b}
l \leq r - 2j + 2i \leq r-2~.
\eeq
\end{enumerate}
%\begin{enumerate}
%\item $j < i$
%\begin{enumerate}
%\item $\kappa = -2 \quad \Rightarrow \quad y \geq 1 \quad \Rightarrow l \leq r$
%\item $\kappa \geq 0 \quad \Rightarrow \quad y \geq 0 \str l \leq r$
%\end{enumerate}
%\item $j=i$
%\begin{enumerate}
%\item $\kappa =-2 \str y \geq 2 \str l \leq r-2$
%\item $\kappa = 0, 1 \str y \geq 1 \str l \leq r-2$
%\item $\kappa \geq 2 \str y \geq 0 \str l \leq r-2$
%\end{enumerate}
%\item $j>i$
%\begin{enumerate}
%\item $\kappa =-2 \str y \geq 3 \str l \leq r-4$
%\item $\kappa \geq 0 \str y \geq j-i \str l \leq r-2(j-i)$
%\end{enumerate}
%\end{enumerate}
\item $q=k+1$
\begin{enumerate}
\item $i=1, 2$

It can be seen from the inequalities (\ref{kappa2}), (\ref{kappa4})
that in this case $\kappa \geq 2$. Using the inequality $i - j \leq 2$
we get from eq.~(\ref{a1})
\beq
\label{bcor3}
l \leq r -2 \quad .
\eeq 
%\begin{enumerate}
%\item $\kappa =-2 \str y \geq 3 \str l \leq r-2$
%\item $\kappa \geq 0 \str y \geq 4 + j-i \geq 2 \str l \leq r -6 - 2(j-i) \leq r-2$
%\end{enumerate}
\item $i =3$

The 
%%%%%%%%%%%%%%%%%%%
general
%%%%%%%%%%%%%%%%%%%
inequality (\ref{kappa3}) leads immediately to
\begin{eqnarray}
\label{bcor*}
\mbox{i.}&&j=0 \,: \quad l \leq r \quad, \\
\mbox{ii.}&&j \geq 1 \,: \quad l \leq r - 1 \quad. \nonumber
\end{eqnarray}
%%%%%%%%%%%%%%%%%%%
\end{enumerate}
\item $q \geq k+2$

Using eq.~(\ref{kappa4}) with the number of pions $p$ given by $p=4q+j-4k-i$
we obtain from eq.~(\ref{a1})
\beq
\label{bcor4}
l \leq r - 6(q-k) +2 +2i - 2j \leq r - 6(q -k) +8 \leq r -4~.
\eeq
\end{enumerate}
The case $i=0$ can be considered analogously:
\begin{enumerate}
\item $ q\leq k-2 \str l \leq r-2$~,
\item $q= k-1$~.
\begin{enumerate}
\item $j \geq 1$

Putting $p = 4 -j$ in eq.~(\ref{kappa2}) we see from eq.~(\ref{a1}), that
\beq
\label{bcor5}
l \leq r-2~.
\eeq
%\begin{enumerate}
%\item $\kappa \geq 0, \; y \geq 0 \str l \leq r-2$
%\item $\kappa =-2, \; y \geq 1 \str  l \leq r-2$
%\end{enumerate}
\item $j=0$

In this case we have $\kappa \geq 2$, as it follows from eq.~(\ref{kappa4}). 
We obtain
\beq
\label{bcor6}
l \leq r~.
\eeq
\end{enumerate}
\item $q=k$

It follows from eqs.~(\ref{kappa1}) 
%%%%%%%%%%%%%%
and (\ref{kappa2}), 
%%%%%%%%%%%%%%
that
\beq
\label{bcor7}
l \leq r - 2~.
\eeq
%\begin{enumerate}
%\item $\kappa = -2 \str y \geq 2 \str l \leq r-2$
%\item $\kappa \geq 0 \str y \geq 1 \str l \leq r-2$
%\end{enumerate}
\item $q \geq k + 1 $

We apply the inequality (\ref{kappa4}) with $p= 4q+j-4k$ to eq.~(\ref{a1})
and obtain
\beq
\label{bcor8}
l \leq r - 6(q -k) +2  \leq r -4~.
\eeq 
\end{enumerate}
\item $E (\lambda^{4k + i}) \lambda^{4k+i} A_l \eta$
\begin{equation}
\label{a2}   
l \leq r~.
\end{equation}
\item $\lambda^{4k+i} A_l \eta H_\kappa \eta$
\begin{equation}
\label{a3}
l = r - \kappa \leq r-2~.
\end{equation}
Here we have used eq.~(\ref{kappa1}).
\item $\lambda^{4k +i} A_l \eta E (\eta)$
\begin{equation}
\label{a4}
l \leq r-2~.
\end{equation}
\item $\lambda^{4k+i} A_{l_1} \eta H_\kappa \lambda^{4q+j} A_{l_2} \eta$
\begin{equation}
\label{a5}
l_1 + l_2 = r - \kappa - 2q -j \leq r -2 \quad .
\end{equation}
%Because $\kappa \geq 0$ for the case $p=0$, $j=1$, it is easy to see, that
%\begin{equation}
%\label{a6}
%l_1, \; l_2 \leq r -\kappa - 6p -2 \leq r-2 \quad .
%\end{equation}

\end{itemize}

%%%%%%%%%%%%%%%%%%%%%%%%%%%%%%%%%%%%%%%%%%%%%%%%%%%%%%%%%%%%%%%%%%%%%%%%%%%%

\setcounter{equation}{0}
\section{Explicit form of the Hamiltonian}
\label{app:H}

In this appendix we give the explicit form of interaction Hamiltonian density,
which we use in this paper. It is based on the effective Lagrangian
given in \cite{ubi1} with an important modification to be discussed below.
\begin{eqnarray}
{\cal H}_1 &=&  \frac{2 g_A}{F_\pi} N^\dagger \fet t \vec \sigma \cdot \cdot
\vec \nabla \fet \pi N           \label{ham1} \\
{\cal H}_2 &=&  \frac{2}{F_\pi^2} N^\dagger \fet t \cdot 
( \fet \pi \times \dot{ \fet \pi} ) N  \label{ham2}\\
&& {} + \frac{1}{2} C_T  
\left( N^\dagger \vec \sigma N \right) \cdot
\left( N^\dagger \vec \sigma N \right)   + \frac{1}{2} C_S
 \left( N^\dagger N \right) \left( N^\dagger N \right)    \label{ham3} \\
%{\cal H}_3 &=& \frac{A_1 '}{F_\pi} \left[ N^\dagger 
%\left( \fet t \vec \sigma \cdot \cdot
%\vec \nabla \fet \pi \right) \vec \nabla^2 N + \vec \nabla^2 N^\dagger \left(
%\fet t \vec \sigma \cdot \cdot \vec \nabla \fet \pi \right) N \right]
%\label{ham4} \\
%&& {} + \frac{A_2 '}{F_\pi}
% \vec \nabla N^\dagger \left( \fet t \vec \sigma \cdot \cdot 
%\vec \nabla \fet \pi \right)
%\cdot \vec \nabla N                        \nonumber \\
{\cal H}_4 &=& C_1 ' \left[  ( N^\dagger \vec \nabla N )^2  +  
( \vec \nabla N^\dagger N )^2  \right]
+ C_2 '  ( N^\dagger \vec \nabla N ) \cdot ( \vec \nabla N^\dagger N ) \nonumber \\
& & {} + C_3 ' ( N^\dagger N ) \left[ N^\dagger \vec \nabla^2 N + \vec \nabla^2
N^\dagger N \right]  \nonumber \\
& & {} + i C_4 '  \left[ ( N^\dagger \vec \nabla N ) \cdot ( \vec \nabla N^\dagger
\times \vec \sigma N ) + ( \vec \nabla N^\dagger N ) \cdot ( N^\dagger \vec \sigma
\times \vec \nabla N ) \right]  \nonumber \\
& & {} + i C_5 '  ( N^\dagger N ) ( \vec \nabla N^\dagger \cdot \vec \sigma \times
\vec \nabla N )  + i C_6 '  ( N^\dagger \vec \sigma N ) 
\cdot ( \vec \nabla N^\dagger
\times \vec \nabla N )  \label{3.194} \label{ham5} \\
& & {} + \left( C_7 '  \delta_{i k} \delta_{j l} 
+ C_8 ' \delta_{i l} \delta_{k j} +
C_9 ' \delta_{i j} \delta_{k l} \right) \nonumber \\
& & {} \times \left[  ( N^\dagger \sigma_k \partial_i N ) 
( N^\dagger \sigma_l \partial_j N ) 
+  ( \partial_i N^\dagger \sigma_k N ) ( \partial_j N^\dagger \sigma_l
 N )  \right]
\nonumber \\
& & {} + \left( C_{10} ' \delta_{i k} \delta_{j l} 
+ C_{11} ' \delta_{i l} \delta_{k j} +
C_{12} ' \delta_{i j} \delta_{k l} \right) 
( N^\dagger \sigma_k \partial_i N ) ( \partial_j N^\dagger \sigma_l N
 )  \nonumber \\
& & {} + \left( \frac{1}{2} C_{13} ' \left( \delta_{i k} \delta_{j l} 
+ \delta_{i l} \delta_{k j}
\right) + C_{14} ' \delta_{i j} \delta_{k l} \right) \nonumber \\
& & {} \times   \left[  ( \partial_i N^\dagger \sigma_k \partial_j N )
 + ( \partial_j N^\dagger
\sigma_k \partial_i N ) \right] ( N^\dagger \sigma_l N )  \nonumber 
\end{eqnarray}
Here we have shown explicitly only the operators ${\cal H}_\kappa$ leading to nonvanishing 
$\eta H_\kappa \eta$, $\lambda^1 H_\kappa \lambda^1$, $\lambda^1 H_\kappa \eta$,
$\lambda^2 H_\kappa \eta$, $\lambda^2 H_\kappa \lambda^1$ and h.~c., which enter eqs.~(\ref{56})--(\ref{60}).
The operators with three or more pion fields are irrelevant.
The symbol '$\cdot \cdot$' means that the appropriate products in
co--ordinate and isospin space have to be taken.

%%%%%%%%%%%%%%%%%%%%%%%%%%%%%
Note  that one has in principle four possible contact terms (two more involving
$\fet t$) in eq.~(\ref{ham3}). However due to Fierz rearrangement  they can be reduced to two.
Further, the 
%%%%%%%%%%%%%%%%%%%%%%%%%%%%%
Hamiltonian used in this paper is always taken 
in normal ordering.

The Hamiltonian given in \cite{ubi1} contains apart from the 
terms enumerated above the two additional interactions
\begin{eqnarray}
{\cal H}_3 &=& \frac{A_1 '}{F_\pi} \left[ N^\dagger 
\left( \fet t \vec \sigma \cdot \cdot
\vec \nabla \fet \pi \right) \vec \nabla^2 N + \vec \nabla^2 N^\dagger \left(
\fet t \vec \sigma \cdot \cdot \vec \nabla \fet \pi \right) N \right]
\label{ham6} \\
&& {} + \frac{A_2 '}{F_\pi}
 \vec \nabla N^\dagger \left( \fet t \vec \sigma \cdot \cdot 
\vec \nabla \fet \pi \right)
\cdot \vec \nabla N~,                        \nonumber 
\end{eqnarray}
which lead to significant contributions to the two--nucleon 
potential at next--to--leading order.
We  now show that both terms (\ref{ham6})
can be completely eliminated from the Hamiltonian by integrating by parts and 
using the nucleons' equation of motion. 
For this let us consider the appropriate terms in the relativistic 
Lagrangian which contains exactly two nucleons, one pion and three
derivatives. One can express them as
\begin{equation}
\label{ham7}
\left( {\bar N}  a^{\mu \nu \rho} 
\left\{ b_1 \lev \partial_\mu \lev \partial_\nu \lev \partial_\rho
+ b_2 \lev \partial_\mu \lev \partial_\nu \pr \partial_\rho 
+ b_3   \lev \partial_\mu \pr \partial_\nu \pr \partial_\rho
+ b_4   \pr \partial_\mu \pr \partial_\nu \pr \partial_\rho \right\}
\fet t N \right) \cdot \fet \pi + {\rm h.c.} \quad .
\end{equation}
Note that one can easily put all other terms with one or more 
derivatives acting on the  pion fields 
into the form (\ref{ham7}) by integrating by parts.  
The most general structure of the quantity $a^{\mu \nu \rho}$  is given by
\begin{eqnarray}
\label{ham8}
a^{\mu \nu \rho} &=& a_1 \gamma^\mu g^{\nu \rho} + a_2 \gamma^\nu g^{\mu \rho}
+ a_3 \gamma^\rho g^{\mu \nu} + a_4 \gamma^\mu \gamma_5 g^{\nu \rho} 
+ a_5 \gamma^\nu  \gamma_5 g^{\mu \rho}
+ a_6 \gamma^\rho \gamma_5 g^{\mu \nu} \\
&& {} + a_7 \epsilon^{\mu \nu \rho \sigma} \gamma_\sigma +
 a_8 \epsilon^{\mu \nu \rho \sigma} \gamma_\sigma \gamma_5~.  \nonumber 
\end{eqnarray}
Now one can see, that all terms with the totally antisymmetric tensor
$\epsilon^{\mu \nu \rho \sigma}$ do not contribute because of the
derivatives commute (i.e. are symmetric under interchange of the indices)
\begin{displaymath}
\partial_\mu \partial_\nu N = \partial_\nu \partial_\mu N~.
\end{displaymath} 
For all remaining terms with the metric tensor $g^{\mu \nu}$
one can make use of the equation of motion of nucleon via
\begin{equation}
\label{ham9}
\left( i \gamma^\mu \partial_\mu - m + \ldots \right) N = 0 
\end{equation}
where the ellipsis represents terms of higher chiral dimension.
%%in order to completely eliminate these terms from the Lagrangian.
We conclude that there are no corresponding terms in the
relativistic Lagrangian. 
Nevertheless one could expect that terms of the type eq.~(\ref{ham6})
arise  in the Lagrangian with fixed coupling constants 
after performing the  nonrelativistic 
expansion for nucleons. In the one--nucleon sector such nonrelativistic
expansion can be carried out explicitely using the 
path integral formulation heavy baryon formalism or by means of
extended reparametrization invariance. One finds 
that one of the terms (\ref{ham6}) indeed enters the Lagrangian representing
a $1 / m^2$ relativistic correction \cite{bkmrev}. 
However when the nucleon mass is counted via (\ref{38}), this term 
is obviously not relevant to the order we are working. 
Only the leading $1 /m$ correction  with exactly two derivatives resulting 
from the Lagrangian of  lowest chiral dimension and 
representing the nucleon kinetic energy is
to be taken into account explicitely.
Our conclusion about the absence of terms (\ref{ham6}) 
in the pion--nucleon Lagrangian agrees
with corresponding result in the one--nucleon sector \cite{ecker}\cite{eckmoj}.
Finally it should be mentioned, that the structure (\ref{ham6})
can not arise by going from the Lagrangian to the corresponding 
Hamiltonian \cite{ubi1}.

%%%%%%%%%%%%%%%% REFS %%%%%%%%%%%%%%%%%%%%%%%%%%%%%%%%%%%%%%%%%%%%%%%%

%%%%\bigskip \bigskip \bigskip

%%%%%%%%%%%%%%%FIGS%%%%%%%%%%%%%%%%%%%%%%%%%%%%%%%%%%%%%%%%%%%%%%%%%%%%%
\newpage

%%%%\end{document}  %%%% intermediate stopx
\section*{Figures}

\vspace{0.5cm}

\begin{figure}[h]
   \vspace{0.9cm}
   \epsfysize=3.2cm
   \centerline{\epsffile{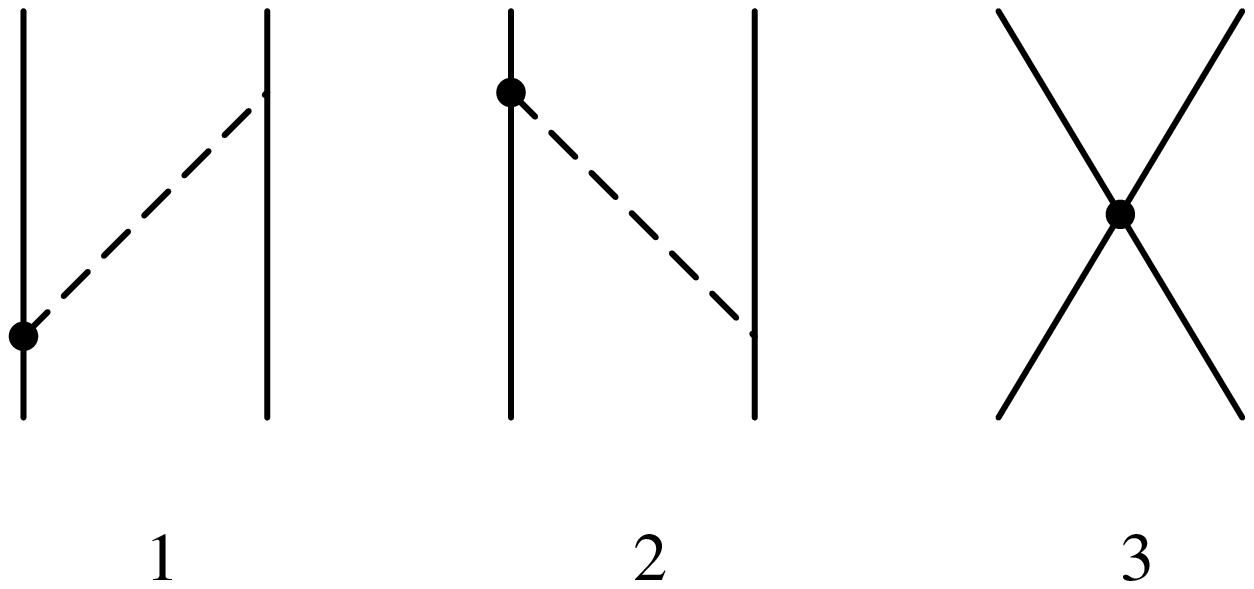}}
   \centerline{\parbox{13cm}{\caption{\label{fig1a}
First corrections to the NN potential in the projection
formalism: One--pion exchange and contact diagrams.
 The heavy dot denotes an insertion from the
next--to--leading order pion--nucleon Lagrangian. Solid and
dashed lines are nucleons and pions, respectively. 
%Graphs which 
%result from the interchange of the two nucleon lines are not shown.
  }}}
\end{figure}
\begin{figure}[bht]
   \vspace{0.4cm}
   \epsfysize=8.7cm
   \centerline{\epsffile{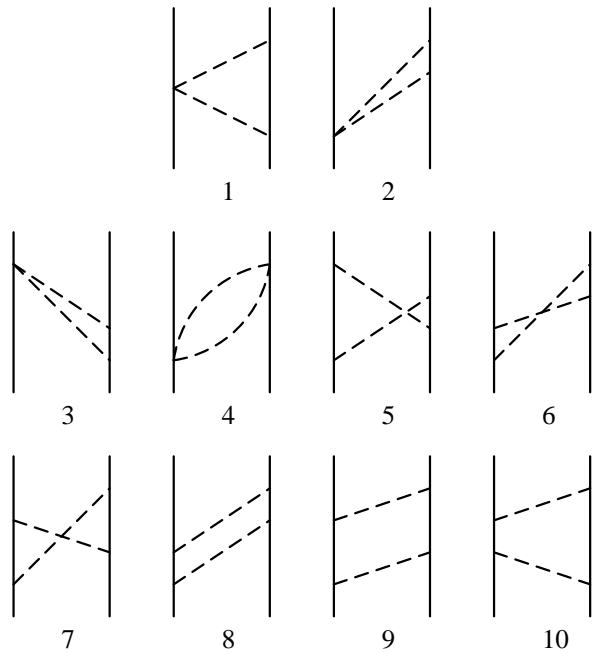}}
   \centerline{\parbox{13cm}{\caption{\label{fig1b}
First corrections to the NN potential in the projection
formalism: Two--pion exchange diagrams. Graphs which 
result from the interchange of the two nucleon lines are not shown.
For notations, see fig.~1.
  }}}
\end{figure}

\begin{figure}[t]
   \vspace{0.5cm}
   \epsfysize=7cm
   \centerline{\epsffile{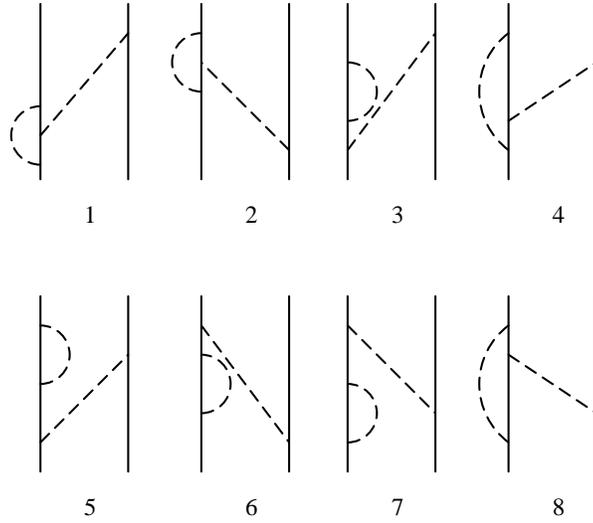}}
   \centerline{\parbox{11cm}{\caption{\label{fig2}
First corrections to the NN potential: Irreducible self--energy and
vertex correction graphs. For notations, see fig.~1.
  }}}
\end{figure}

\begin{figure}[b]
   \vspace{0.5cm}
   \epsfysize=7cm
   \centerline{\epsffile{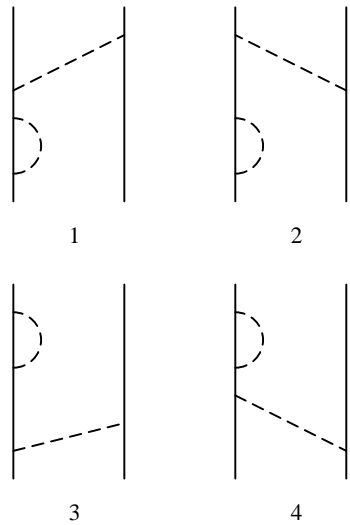}}
   \centerline{\parbox{11cm}{\caption{\label{fig3}
First corrections to the NN potential: Reducible self--energy
graphs. For notations, see fig.~1.
  }}}
\end{figure}

\newpage

\begin{figure}[t]
   \vspace{0.5cm}
   \epsfysize=5cm
%%%   \centerline{\epsffile{lee3.ps}}
   \centerline{\epsffile{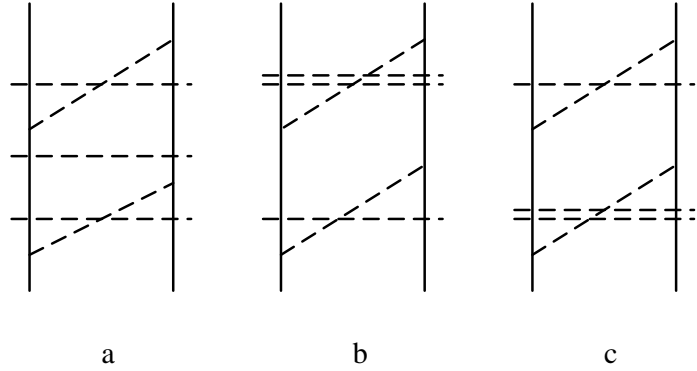}}
   \centerline{\parbox{11cm}{\caption{\label{fig4}
Reducible graphs. In (a), a truly reducible diagram is shown.
In the projection formalism, one has graphs like
(b) and (c). These correspond to diagram~9 in fig.~2. The horizontal
dashed lines count the free energies of the particles cut.
  }}}
\end{figure}

\begin{figure}[b]
   \vspace{0.5cm}
   \epsfysize=7cm
%%%   \centerline{\epsffile{lee3.ps}}
   \centerline{\epsffile{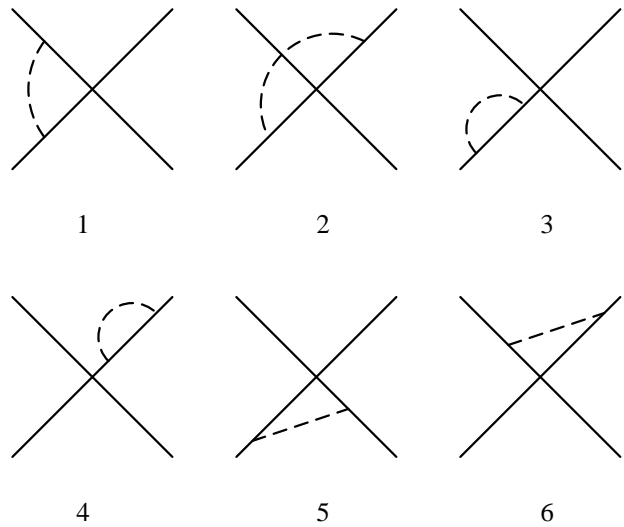}}
   \centerline{\parbox{11cm}{\caption{\label{fig5}
One--loop corrections to the four--nucleon
contact interaction.  For notations, see fig.~1.
  }}}
\end{figure}

\newpage

\begin{figure}[t]
   \vspace{0.5cm}
   \epsfysize=7cm
%%%   \centerline{\epsffile{lee3.ps}}
   \centerline{\epsffile{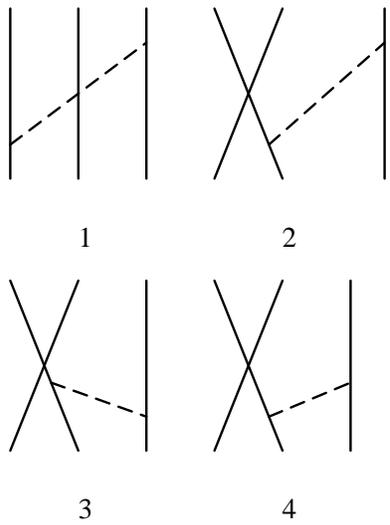}}
   \centerline{\parbox{11cm}{\caption{\label{fig6a}
Leading diagrams contributing to the three--nucleon force
that vanish. (1) is a representative for all graphs involving the
$\pi\pi NN$--vertex. (2,3,4) all involve one four--fermion
   interaction.
For notations, see fig.~1.
  }}}
\end{figure}

\begin{figure}[b]
   \vspace{0.5cm}
   \epsfysize=7cm
%%%   \centerline{\epsffile{lee3.ps}}
   \centerline{\epsffile{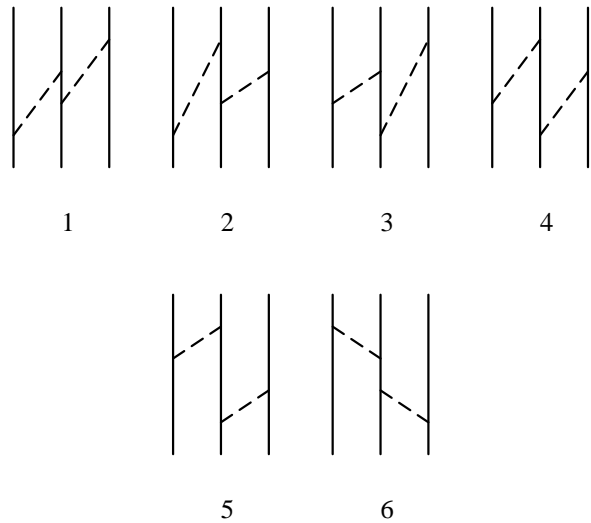}}
   \centerline{\parbox{11cm}{\caption{\label{fig6}
Leading diagrams contributing to the three--nucleon force.
{}From the reducible graphs, only one
representative is shown (as explained in fig.~5).
For notations, see fig.~1.  
  }}}
\end{figure}

\end{document}